\documentclass[review,authoryear,12pt]{elsarticle}
\usepackage{graphicx} % for including graphics
\usepackage{xcolor}
\usepackage{soul}
\usepackage{amssymb}
\usepackage{multirow}
\usepackage{xcolor}
\usepackage{siunitx} %% micro meter

\usepackage{natbib}
\setcitestyle{compress,square,numbers,comma,sort}

\journal{Ultrasound in Medicine and Biology}

\begin{document}

\begin{frontmatter}

%% Title


\title{Quantitative Ultrasound for Periodontal Soft Tissue Characterization}

%% Authors and addresses/affiliations

%% use the corref command within \author for corresponding author footnotes; note the corresponding author can be any of the authors, but one author must be designated; here the third author has been arbitrarily designated as the corresponding author as an example.



\author[Affil1]{Daria Poul \corref{cor1}}
\author[Affil1]{Ankita Samal}
\author[Affil2]{Amanda Rodriguez Betancourt}
\author[Affil1]{Carole Quesada}
\author[Affil2]{Hsun-Liang Chan}
\author[Affil1,Affil3]{Oliver D. Kripfgans \corref{cor2}}
\address[Affil1]{Department of Radiology, University of Michigan School of Medicine, Ann Arbor, Michigan, United States}
\address[Affil2]{Department of Periodontics and Oral Medicine, University of Michigan School of Medicine, Ann Arbor,  Michigan,  United States}
\address[Affil3]{Department of Biomedical Engineering, University of Michigan, Ann Arbor, Michigan,  United States}
\cortext[cor1]{Corresponding Author 1: ssheykho@umich.edu} \cortext[cor2]{Corresponding Author 2: greentom@umich.edu}

\begin{abstract}

\noindent \textbf{Objective} 
\newline Periodontal diseases affect 45.9\% of adults of age $\geq 30$ years in the United States. Current clinical diagnostic methods are invasive, subjective and qualitative. Quantitative Ultrasound (QUS) has shown promising results in noninvasive characterization of various soft tissues; however, it has not been used in clinical periodontics.\newline
\textbf{Methods}
\newline \noindent Here, we investigated QUS analysis of two periodontal soft tissues (alveolar mucosa and gingiva) \textit{in vivo}. Study cohort included 10 swine scanned at four oral quadrants, resulting in 40 scans. The two-parameter Burr and Nakagami were employed for QUS-based speckle modeling. Parametric imaging of these parameters was also created using an optimal estimation window size (WS) in a separate phantom study. 
\newline
\textbf{Results}
\newline \noindent Phantom results suggested a WS of 10 wavelengths as the reasonable estimation kernel. The Burr power-law parameter and Nakagami shape factor were higher in gingiva than alveolar mucosa while the Burr and Nakagami scale factors were lower in gingiva. The difference between the two tissue types was statistically significant (p$<$0.0001). Linear classifications of these two tissue types using a 2D parameter space of the Burr and Nakagami models resulted in a segmentation accuracy of 93.51\% and 90.91\%, respectively. Findings from histology-stained images showed that gingiva and alveolar mucosa had distinct underlying structure with gingiva showing a denser stain.\newline 
\textbf{Conclusion}
\newline \noindent QUS results suggested that the gingiva and alveolar mucosa were differentiable using the Burr and Nakagami parameters. We propose that QUS holds promising potential for characterization of periodontal soft tissues. QUS could become an objective and quantitative diagnostic tool for periodontology and implant dentistry to improve dental healthcare.\newline
\end{abstract}

\end{frontmatter}

%% Do not remove the page break here.
\pagebreak

% \linenumbers

%% MAIN TEXT INSTRUCTIONS

%% For all sections, subsections, and subsubsections, use the '*' to remove numbering, as demonstrated below.

%% Commands for figures and tables should not be included in the main body of the submitted version of this file (e.g. the figure and tabular environments).  Figure captions should be listed in this file, as shown below.  Tables and Table captions should be listed as a separate section at the end of this file, as shown below.  Many authors may wish to include figures and tables within the main text of their document will preparing their manuscript.  This may be done, however please comment out any of the lines prior to submission.

%% Because the Elsevier editorial process does not allow for the figure and tabular environments in the submitted document, you will be unable to use autonumbering (i.e. \label and \ref) for figures and tables. 

%%  If long equations are used in the document, authors should use a two column format to make sure that the equations will break at approximately the right places.  To do this, replace the class option 'review' with the following two class options '3p' and 'twocolumn'.  Keep in mind that the column width produced in '3p' is slightly narrower than the final printer version.  After inserting the appropriate line breaks in your equation, change the '3p' option back to 'review'.

%% For citations, use the commands \citep and \citet

%% BEGIN MAIN TEXT

%%%%%%%%%%% INTRODUCTION
\section{Introduction}
\label{intro}
Periodontal (gum) diseases are reported to affect nearly half (45.9\%) of the adult population aged $\geq 30$ years in the United States \cite{RN23}. These diseases concern various oral soft tissues that support and surround teeth such as marginal (free) gingiva, attached gingiva and, alveolar mucosa, which are illustrated in \textbf{Figure \ref{anatomy}} for a swine model. The most prevalent periodontal diseases affecting these tissues are periodontitis and gingivitis which are a continuum of inflammatory diseases. Periodontitis is initiated by bacteria infection, potentiated by inflammation, resulting in periodontal attachment loss of soft tissues from bone as well as causing actual bone loss. As periodontitis progresses, it could cause tooth loss. It is also related to systemic diseases, such as cardiovascular diseases and diabetes. On the other hand, gingivitis is considered reversible, involving gingival inflammation without clinical signs of bone loss \cite{RN24}. If oral diseases, affecting both soft and hard tissues, are not addressed at early stages, those could impose immense pain as well as excessive economic burdens on the population, (it is reported to have caused direct and indirect burdens of as high as \$154.06B in the United States in 2018 \cite{RN25}.\newline
Among diagnostic modalities in dentistry for clinical assessments of soft tissue is bleeding on probing (BOP). BOP is an invasive method in which a probe with ruler markings of 1 mm increments is inserted and gently pushed into the pocket/sulcus between the crown and the marginal (free) gingiva (see \textbf{Figure \ref{anatomy}}). Probing depth as well as potential bleeding are frequently recorded at office visits as a surrogate for gingival inflammation and are a part of standard of care as BOP is a sign of periodontal inflammation. Normal probing depth is equal or less than 3 mm, beyond which attachment/bone loss around a tooth is suspected. Apparently, BOP has several limitations: it is invasive, often time exerting unpleasant experiences on patients. Also, BOP is subjective, as penetration force/insertion angle of the probe and tissue texture can vary the readings significantly. Also, probing depth is insensitive in the sense that it is only able to differentiate increments of 1 mm and small-scale penetration depths bare reading errors. Additionally, common variations in gingival thickness induced by different biotypes in different patients could  increase the complexity of obtaining an objective assessment of inflammation by BOP \cite{RN11}. This method is qualitative as it describes either no, slight, or profuse/spontaneous bleeding. BOP is a measure of tissue destruction at a late and already irreversible stage \cite{RN40}. It is noteworthy that although lack of BOP observation is a strong indication of negative inflammation, BOP observations do not necessarily indicate the existence of an underlying inflammation \cite{RN41}.  Another traditional diagnostic method in dentistry is the visual observation which suffers from some of limitations as those listed for BOP. For example, swollen and erythematous tissue is indicative of periodontal inflammation. However, pigmented and thick tissue can mask these cardinal signs. Therefore, it is crucial to investigate non-invasive diagnostic modalities for objective and quantitative characterization of oral soft tissues from clinical workflow aspects as well as for improving public health and alleviating the associated financial burden. \newline
Towards this clinical goal, an imaging modality with promising potentials for oral soft tissue characterization is ultrasound (US) imaging. As a non-invasive, non-ionizing, real-time, inexpensive, and well-established modality, US has been employed for imaging and characterization of various biological soft tissues such as liver, thyroid, muscle, etc. at differing depths and image resolutions  \cite{RN44,RN14,RN45}. In dentistry, US B-mode (brightness-mode) imaging has been employed for lesion detection, measuring gingiva thickness and to delineate the surface of hard tissues (bone/crown) \cite{RN13,RN16,RN17,RN18}. Moreover, ultrasound-based elasticity estimations of oral soft tissues have been investigated in different studies \cite{RN15, RN26}. Ultrasound imaging offers information beyond B-mode imaging. One important aspect to employ ultrasound imaging is quantitative ultrasound (QUS). In QUS analysis of tissues, the goal is to find quantitative parameters from uncompressed raw ultrasound scan data that can be linked to some measure of underlying structure of tissues, which could offer clinical potentials for tissue characterization \cite{RN14,RN19,RN36}. While B-mode images provide information about landmark anatomical structures of tissues, it fails to provide information and contrast of the underlying soft tissue structure. QUS parameters could add more information to B-mode images, represented as a parametric image overlay. Although QUS analysis has been extensively applied to characterize various biological tissues \cite{RN48, RN49, RN50, RN51, RN4}, it has never been applied in clinical periodontics for soft tissue characterization  and there are only few studies with limited analysis involving QUS for periodontal soft tissues. For example, in the study reported in \cite{RN32},  US B-scan echogenicity in layers of oral soft tissues were compared using a measure of echo levels. Nevertheless, the employed method deviates from standard techniques of analyzing US image echogenicity parameters quantitatively. Potential application of QUS in periodontology includes tissue characterization, i.e., gingiva versus a. mucosa, tracking inflammation progress, automatic lesion detection, tissue healing quantification, etc. QUS could potentially complement BOP and other diagnostic tools for an improved and more accurate characterization of periodontal tissues.\newline
One class of QUS analysis in medical imaging of tissues is focused on modeling first order statistics of ultrasound speckle. Speckles are granular (grey) textures observed in B-mode tissue images \cite {RN53}. Speckles result from the interference of backscattered waves echoed back from various tissue scatterers close to each other during ultrasound pulse-echo imaging. Although speckles have a negative impact on B-mode image quality and are filtered out for US image representation, speckle patterns could incorporate information about underlying tissue structure and thus, could have clinical significance. Modeling ultrasound speckle statistics may result in deriving quantitative parameters that can be correlated to tissue pathology not visible on B-mode images. A number of well-establish distributions for speckle modeling in QUS include Rayleigh \cite{RN46, RN47}, Homodyned-K \cite{RN38,RN37,RN39}, Nakagami \cite{RN22,RN21,RN20}, and more recently the Burr model \cite{RN9,RN5,RN1}. All these distributions have been widely used for QUS-based tissue characterization.
However, to the best of our knowledge, characterization of periodontal soft tissues using  speckle modeling and these well-established distributions have not been reported in the literature. Here, we aim at characterizing periodontal soft tissue  by investigating US speckle statistics using the Burr and Nakagami models in an \textit{in vivo} animal study on swine oral tissues. Moreover, we present parametric imaging of these QUS parameters as an additional information to that of B-mode images. Additionally,  histology images of swine oral tissues were acquired using Masson’s Trichrome and hematoxylin–eosin (H\&E) stains with a 20x magnification microscopy imaging to compare tissue structures with  QUS analysis.  Gingival and alveolar mucosal tissues were compared. QUS-based parametric imaging may have potential to be used as an augmented tool to current imaging modalities in dentistry such as cone beam computed tomography (CBCT) to further aid oral surgeons and to provide diagnostic value to clinical assessment in oral examinations. \newline
Before we delve into the QUS analysis, we will first present a concise overview of oral soft tissue anatomy to introduce necessary information and terms used within the rest of the paper. 
\subsection{\textbf{Oral Soft Tissues: Anatomy}}
\begin{figure}[b]
\centering
\includegraphics[width=1\textwidth,trim={1.2in 0.9in 1.4in 3in}]{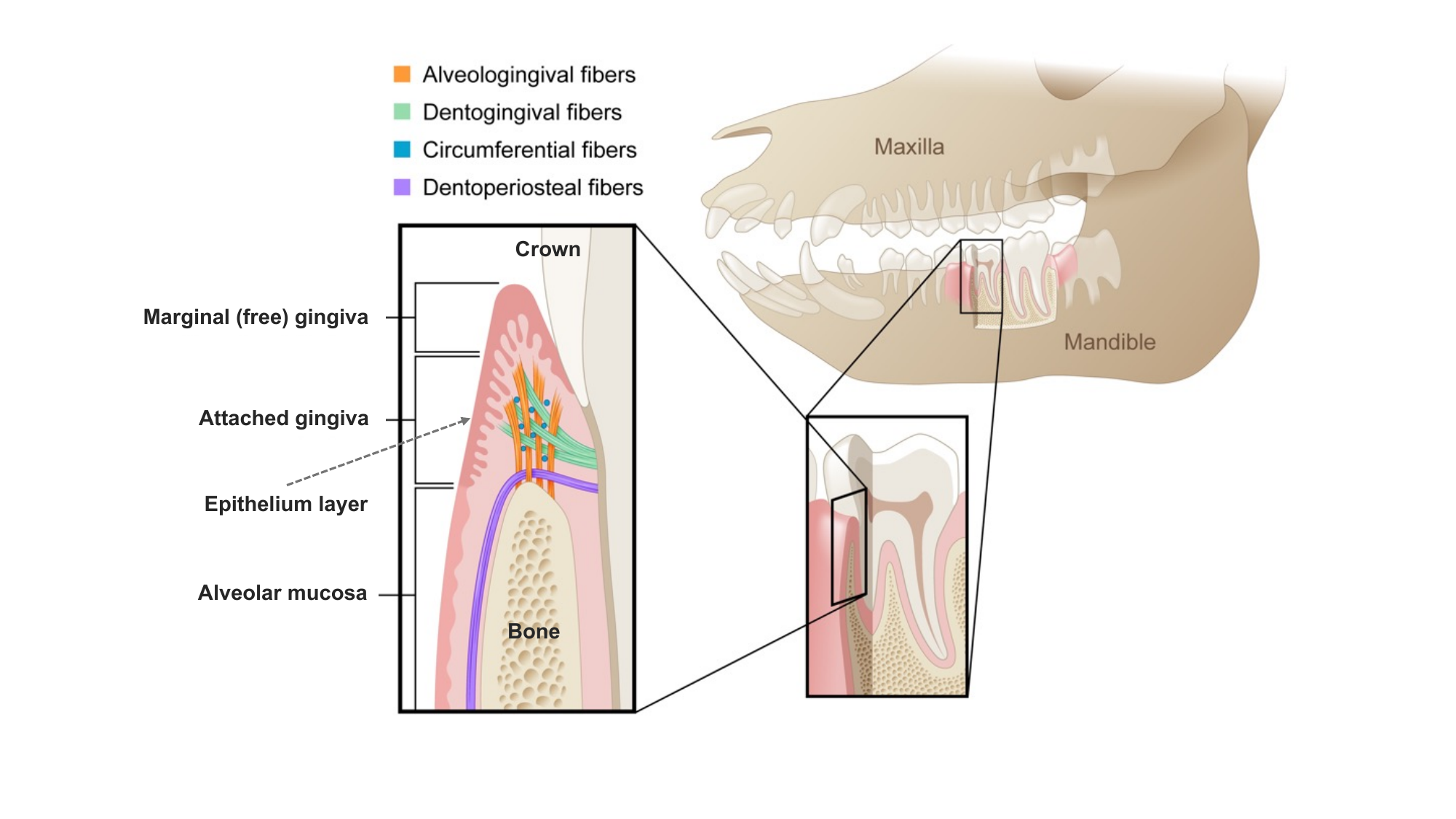} % [left lower right upper]change 'example-image' to the filename of your image 
\caption{Anatomical structure of swine periodontal tissues. Several types of fibers within the gingival tissues provide it with mechanical support during mastication (chewing).}
\label{anatomy}
\end{figure}

Oral soft tissues are comprised of different components, such as gingiva, alveolar mucosa, buccal mucosa, and epithelium, each with different physiological properties suitable for a particular function during the mastication (chewing) process. The gingiva (G) and alveolar mucosa (M) are located closely in the proximity of teeth (crowns); however, the two types of tissues are distinct and have different ultrastructure. They are considered two important components of oral soft tissues that have drawn significant clinical studies due to the frequent occurrence of dental issues associated with qualitative and/or quantitative changes in gingiva and alveolar mucosa tissues \cite{RN27}. These tissues along with other structures of hard and soft tissues are illustrated in \textbf{Figure \ref{anatomy}} in a swine model. 
\subsubsection*{Gingiva}
Gingiva is a dense fibrous connective tissue having load-bearing intracellular layers and its primary role is to protect the root and alveolar bone from deformation and degradation. Connective tissues within the gingiva are mostly comprised of collagen fibers organized in different patterns: (i) thick collagen fibers arranged densely, and (ii) short and thin collagen fibers arranged sparsely along with fine-sized network (reticular) structures of fibers as well as some diffuse collagen fiber patterns \cite{RN30}. Gingiva comprises various fiber arrangements as also illustrated in \textbf{Figure \ref{anatomy}}, which includes: alveologingival fibers (originating from bone to epithelium), dentogingival fibers (stretched from tooth root to gingival region), circumferential fibers (encompassing the tooth), and dentoperiosteal fibers (from the tooth root to the bone). In terms of vasculature, gingiva has dense capillary vessels ($\geq15$ \textmu m in diameter) that are mostly perpendicular to the gingival surface and lack connective vessels, with sparse large vessels at higher depth \cite{RN31}. 
 The gingiva consists of two main parts: the free gingiva, which wraps around the tooth and is free (not attached) from one side and, the attached gingiva which is attached to the free gingiva from one side and is firmly connected to the alveolar bone on the other side. Among periodontal soft tissues, the gingiva has relatively higher exposure to the external mechanical forces (cyclic and non-cyclic) applied from mastication compared to other oral soft tissues and thus, has a stiffer nature \cite{RN28}. The gingiva is covered by an additional highly keratinized layer called epithelium (E) that forms a biological seal around the gingiva, lowering penetration of some substances into the oral soft tissues beneath it \cite{RN29,RN28}.
\subsubsection*{Alveolar Mucosa} 
Alveolar mucosa is a membrane that lines the bone. Unlike gingiva, alveolar mucosa is less exposed to abrasive forces and is mainly non-keratinized. It has a higher level of elastic fibers which makes the alveolar mucosa more elastic compared to the gingiva. The gingiva contains higher collagen fiber levels cross-linked and possesses some resistance to tensile loads. These elastic fibers tend to make alveolar mucosa return to the resting state when being extended. Moreover, it is distinguished by a higher  blood vessel supply, ranging from capillaries to larger ones and appears pinkish compared to the gingiva's brighter white-pink color \cite{RN33}. The interstitial fluid within the vasculature provides cushioning when tissue is under large masticatory loads.

%%%%%%%%%%% MATERIALS AND METHODS
\section{Theory}
\label{Theo}
In QUS,  the first order speckle statistics is the probability distribution of the envelope of ultrasound echo amplitudes. Modeling speckle statistics could provide information about scatterer structure within tissues.  This section provides the theoretical background for modeling first order ultrasound speckle statistics and also QUS parametric imaging using Burr and Nakagami models.      
\subsection{\textbf{The Burr Model for Speckle Statistics Modeling}}
Recently, a new framework has been proposed to model the first order statistics of ultrasound echo amplitude from tissue backscattering which is based on a key assumption that scatterers within tissues are multi-scale fractal and their number density follows a power-law distributions with the characteristic size of scatterers ($b$ as the key power-law parameter related to scatterers' density). The mathematics under this framework resulted in the Burr distributions for describing the first order statistics of ultrasound echo amplitude, which was the first application of the Burr model within the area of medical imaging \cite{RN9,RN5,RN1}. The Burr distribution was first derived in the 1940s without any implications to the field of ultrasound medical imaging \cite{RN52}. 
This framework was initially employed to describe ultrasound speckle statistics from \textit {in vivo} livers in normal and abnormal conditions \cite{RN3,RN4} as well as for describing the speckle statistics from a set of simulated scattering structures in the form of cylindrical and spherical scatterers in which number densities of multi-scale scatterers followed a power-law distribution with radii \cite{RN1,RN2,RN6}. The results from these studies showed that the Burr distribution successfully and efficiently described US speckle statistics. Later, the Burr distribution was employed in optical coherence tomography (OCT) scans and it was reported that the Burr distribution shows promising results in modeling speckle statistics in OCT scans \cite{RN8,RN35}.
Under this framework, the histogram of backscattered echo amplitudes (A) can be modeled as a probability density function (PDF), denoted as P(A) in equation ~\ref{eq:1} with two underlying parameters: the key power-law parameter $b$ and a scale factor $l$, as  following:
\begin{equation} 
    \label{eq:1}
    P(A)=\frac{2A(b-1)}{l^2[(\frac{A}{l})^2+1)]^b}
\end{equation}
Burr \textit{b} is associated with the number density of scatterers and increases with that while Burr \textit{l} is related to the echo amplitude and is elevated in tissues with higher echogenicity. These parameters provide additional information about the tissue’s underlying structure compared to the grey-scale  B-scan alone. Burr \textit{b} and \textit{l} have shown potential in characterizing liver tissues in  fibrotic  vs. normal conditions \cite{RN4, RN3}. 
To estimate the Burr parameters from the tissue backscatter within a selected region of interest (ROI), we can fit the echo envelope PDF of the speckle data to equation ~\ref{eq:1} and derive the underlying parameters. Also, the Burr parameters can be estimated locally from the local statistics of speckle data using a sliding window approach where the ROI is swept by a small kernel and a local estimation map of the Burr parameters is calculated. To do so, we could utilize relationships between one or multiple statistical moments of the echo amplitude and the Burr parameters to find a system of two equations with two unknown parameters \cite{RN7}. One statistic to employ is the first moment of the echo amplitude, i.e. mean: denoted as $E[A]$ and reported in equation~\ref{eq:2}. Other statistics could be the ratio of the square of the first moment of the echo amplitude, $(E[A])^2$, to the first moment of the echo intensity, $E[A^2]$, as shown in equation~\ref{eq:3}.
While equation~\ref{eq:2} shows that the first moment depends on both \textit{b} and \textit{l}, per equation ~\ref{eq:3} the moments’ ratio depends \textit{only} on one single parameter, (\textit{b}). To estimate these two parameters, first \textit{b} is obtained from Equation~\ref{eq:3} and then, it is plugged into equation~\ref{eq:2} to estimate \textit{l}. The Burr scale factor, \textit{l}, in equation ~\ref{eq:2} is constrained by the finite value of \textit{b} (measure of the scatterer number density, from equation ~\ref{eq:3}) and by the value of the first moment of the amplitude (mean), \textit{E[A]}. Thus, \textit{l} does not diverge in practice due to the finite nature of both the backscatter amplitude and the \textit{b} parameter. Using these two equations, one can obtain local estimations of the Burr parameters.
\begin{equation}
    \label{eq:2}
    E[A]=\frac{(b-1)l\sqrt{\pi}\Gamma(b-\frac{3}{2})}{2\Gamma(b)}
\end{equation}
\begin{equation}
    \label{eq:3}
    \frac{(E[A])^2}{E[A^2]}=\frac{(b-2)\pi(\Gamma(b-\frac{3}{2}))^2}{4(\Gamma(b-1))^2}
\end{equation}
\subsection{\textbf{Nakagami Distribution}}
The probability distribution of the echo amplitude envelope from the two-parameter Nakagami distribution is modeled according to equation~\ref{eq:4}  where $m$ is the Nakagami shape parameter and $\Omega$ is the Nakagami scale factor. These two parameters are estimated statistically from equation~\ref{eq:5}  and equation~\ref{eq:6}. It is noted that the Nakagami $m$ parameter determines the form of the speckle statistics PDF: if $m<1$, it is pre-Rayleigh and a heavy-tailed distribution, if $m=1$, it corresponds to Rayleigh behavior and for $m>1$, it demonstrates a post-Rayleigh distribution \cite{RN20}. 
The Nakagami scale factor $\Omega$ represents the total intensity of the backscattered echo within the region under analysis.
\begin{equation}
    \label{eq:4}
    f(r)=\frac{2m^mr^{2m-1}}{\Gamma(m)\Omega^m}e^{-\frac{m}{\Omega}r^2}U(r)
\end{equation}
\begin{equation}
    \label{eq:5}
    m=\frac{(E[A^2])^2}{E[A^2-E[A^2]]^2}
\end{equation}
\begin{equation}
    \label{eq:6}
    \Omega=E[A^2]
\end{equation}

\noindent In these equations, $f(r)$ is the Nakagami density function and $U(r)$ is the unit step function. The Nakagami shape factor \textit{m} determines the form of the probability distribution function of speckle data and is related to the scatterer concentration while the Nakagami scale factor $\Omega$ quantifies the backscatter intensity \cite{RN20}. These parameters have been used in various studies for tissue characterization. 
For example, in an animal model study reported in \cite{RN21}, it was shown that the Nakagami shape factor increased with increase in the liver fibrosis score in rats. This was postulated to result from the increase in the concentration (density) of local scatterers and also the appearance of periodic structures or clustering of scatterers in tissues, leading to a post-Rayleigh statistical distribution and higher \textit{m}.

%%%%%%%%%%% MATERIALS AND METHODS
\section{Materials and Methods}
\label{MaM}
\subsection{\textbf{Study Design for Intraoral US Imaging of Periodontal Soft Tissues}}
For the QUS investigations of periodontal soft tissues \textit{in vivo}, there are some challenges associated with intraoral US imaging itself which require an intricate study design and customized scanning setup to tackle them \cite{RN12}. Two most important ones are dealing with the intraoral scanning of inherently small-sized periodontal soft tissues and also the mixed presence of hard and soft tissues. In addition, in intraoral scanning, the buccal (cheek) region imposes a less accommodating condition to foreign objects such as an ultrasound transducer. This limits the capability of freely placing and maneuvering the transducer when scanning the tissue. To address these challenges, we benefited from using a, recently introduced, toothbrush-sized  high-frequency US transducer \cite{RN43}. Its  physical design and ability to provide high-resolution images was made for the purpose of intraoral imaging \cite{RN12} through a collaboration of clinicians and scientists (co-authors H-L.C. and O.D.K) at the University of Michigan (Ann Arbor, MI) and the Mindray Innovation Center (Mindray Inc., San Jose, CA). This US transducer is shown in \textbf{Figure \ref{tdr}} \textbf{(a)} and \textbf{(b)}, with its in situ placement in \textbf{Figure \ref{tdr}} \textbf{(c)}. The center frequency of the transducer is 18 MHz, the imaging depth and the lateral field of view commonly used are approximately 15 mm and 13 mm, respectively and the transducer elevational focus is at a depth of 8 mm. The useable bandwidth of this array is 8 to 30 MHz. The system was operated in a single transmit and receive frequency mode for the QUS data acquisition in this study. The pre-clinical images (not IQ data) were obtained in the standard second harmonic imaging mode (12 MHz transmit, 24 MHz receive) with additional spatial compounding to reduce clutter and for obtaining high-resolution ultrasound images. In order to acquire high-quality gingival and mucosal scans, a standoff gel pad was placed onto the aperture surface to shift the tissues of interest to the focal region of the transducer, as shown in \textbf{Figure \ref{tdr}} \textbf{(d)}. Having in mind the anatomical structure of oral tissues shown in \textbf{Figure \ref{tdr}} \textbf{(e)}, a sample of US scan of periodontal soft tissues in a swine model is presented in \textbf{Figure \ref{tdr}} \textbf{(f)}.  Regions of the gingiva, mucosa, epithelium, muscle as well as crown and bone are annotated to help distinguish regions of interest for QUS analysis. It is noted that in the B-scan presented in \textbf{(f)}, the transducer is located on the right edge of the image.
The B-scan sample shows that the epithelium appears as a hyper-echoic (bright) region, indicating a higher US backscatter intensity from this region. This higher scattering phenomenon may result from contributions of dense keratinized tissue (higher impedance). Also, the normal orientation of the stratified epithelium layer with respect to the transmitted pulse from the US probe may produce the strongest reflection compared to other angles. \newline
For an accurate QUS analysis of periodontal soft tissues, two experimental studies were designed here: the first experiment focused on scanning and analyzing homogeneous custom-designed phantoms without any macro inclusions (CIRC Inc., Virginia, USA), and the second part focused on investigating \textit{in vivo} periodontal soft tissues of swine. The phantom study aimed at finding the optimal window (kernel) size for accurate and robust local QUS parameter estimations. Specifically, it was meant to minimize variability caused by an overly small kernel size and concurrently, to avoid excessively large kernel sizes that result in loss of spatial resolution in QUS parameter estimations.\newline
Swine models were selected for this study due to the resemblance of human and swine oral soft tissues from histology and morphology aspects \cite{RN34}. Animals (N=10) were obtained from Sinclair Bio Resources (Auxvasse, MO, USA) under a study protocol approved by the Institutional Animal Care and Use Committee at the University of Michigan (PRO00010333). Four males and six females were investigated. Intraoral US scanning was performed at the mid-facial location of all four first molars (M1), which we refer to by combinations of L (left) / R (right) and mandible (MAND) / maxilla (MAX). It is noted that a small notch was manually created on all teeth as a landmark visible on ultrasound images. 
\subsection{\textbf{US Data Acquisition and Parametric Imaging}}
\begin{figure}[b]
\centering
\includegraphics[scale=0.5, trim={2.2in 0.5in 0.4in 0in}] {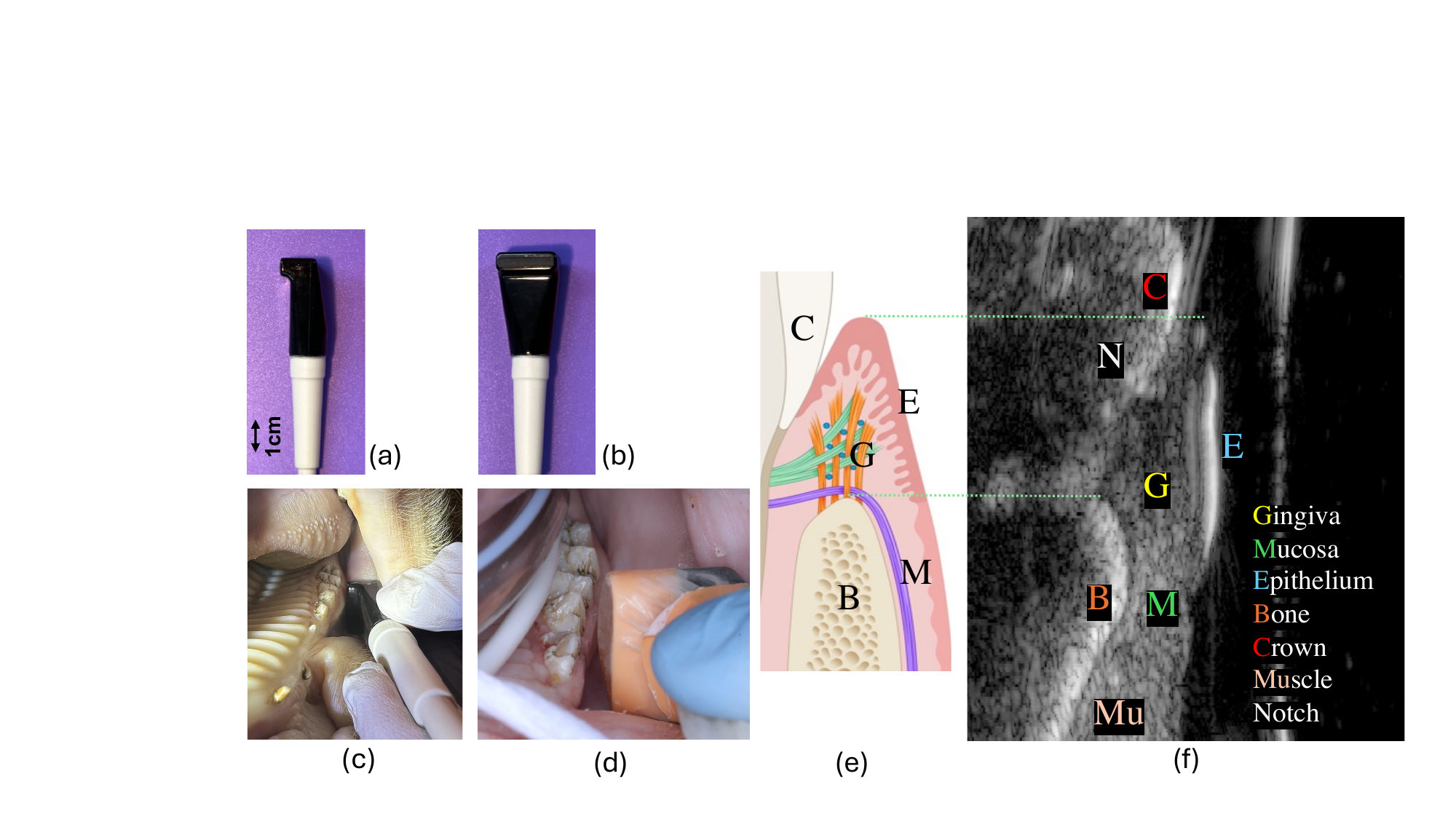}% change 'example-image' to the filename of your image
\caption{\textbf{(a)} and \textbf{(b)} High-frequency toothbrush-sized ultrasound transducer for the intraoral scan, \textbf{(c)} the transducer positioning for mid-facial imaging of a molar tooth within the transverse plane in a swine, \textbf{(d)} zoomed-in view of the transducer with the standoff gel pad, \textbf{(e)} an illustration of oral soft tissues anatomy as a general reference for understanding B-scan structure in \textbf{(f)} for a swine case. Important anatomical structures of periodontal hard and soft tissues are annotated in both \textbf{(e)} and \textbf{(f)}.}
\label{tdr}
\end{figure}

Ultrasound imaging data for both phantom and swine studies were acquired by employing a clinical Mindray ultrasound imaging system (ZS3, Mindray Innovation Center, San Jose, CA, USA) equipped with a high-frequency linear array transducer, now commercially available (L30-8, ZS3, Mindray Inc., San Jose, CA) introduced earlier in \textbf{Figure \ref{tdr}} \textbf{(a)} and \textbf{(b)}. For ultrasound data acquisition, the scanner employs the Zone Sonography\textsuperscript{TM} Technique, proprietary to Mindray Inc., in which a broad-spectrum ultrasound pulse that is transmitted through individual zones to utilize synthetic aperture imaging and potentially cover the field of view in fewer transmit/receive cycles \cite{RN55}. This is contrary to the traditional systems where line-by-line data acquisition only allows for a single focus per transmit thus limiting the imaging depth of field, whereas zone transmits perform only lateral focusing on receive. The raw RF-data were demodulated, and the envelope detected using a Hilbert transform. B-mode images were reconstructed using the logarithmically compressed envelope data within a dynamic range of 70 dB. 
The Burr model parameters ($b$ and $l$), and the Nakagami model parameters ($m$ and $\Omega$) were locally estimated in user-defined ROIs within the gingiva and the mucosa for each swine to provide additional information about tissue structure using an estimation kernel approach. The sliding kernel moved across ROIs (along horizontal and vertical directions) with an overlap ratio of 70\%. Parameter estimations derived from the kernel at each location within the ROI was assigned to the center of the  kernel. Model parameter estimations were mapped as colored parametric images overlaid on the B-mode images. The optimal kernel (window) size for tissue parametric imaging was obtained from the phantom study. \newline
It is noted that parameter estimations were performed in MATLAB R2023 (MathWorks, Inc., Natick, Massachusetts). To solve the nonlinear equation for estimating the Burr power-law \textit{b}, equation~\ref{eq:3} is rearranged by shifting all terms to the left size and setting the equation to zero. Then, the zeros of the resulting equation are sought for numerically using a nonlinear solver and a reasonable random initial guess. For estimating the remaining model parameters, associated equations are used directly, as outlined in the Theory section.

\subsubsection{Optimal Window Size}
In statistical estimations of the Burr and Nakagami parameters for parametric images, there is a trade-off between the spatial resolution of local parameter estimation and statistical variability in estimated parameters. If the window size (WS) is too small, the speckle within the kernel provides unrobust estimations, i.e., the parameters don’t converge. On the other hand, the size of respective tissues under study imposes limitations on the size of the associated ROI and consequenty, how large the window can be. The gingival and alveolar mucosal tissues are inherently small, often about 2 mm thick, which limits the ROI size for the QUS analysis.  This trade-off was investigated by evaluating model parameters as a function of WS. Phantom scan envelope data were used to create parametric images of rectangular ROIs (5 mm lateral by 2.5 mm axial) located in the center of the field of view at four different imaging scans. Overall results of these four scans are presented as a boxplot for each WS. In the phantom study, we tested kernels with various sizes each being an integer multiple of the imaging wavelength (64.2 µm based on 1,540 m/s and 24 MHz receive frequency), as the size unit for reference. For this analysis, we followed a standard procedure in finding the optimal window size \cite{RN51, RN20}. Nine different window sizes from 2 to 18 wavelengths with an interval of 2 were tested. Data processing was performed using MATLAB (R2023a, MathWorks Inc., Natick MA, USA).
\subsubsection{ROI Selection Criteria for Outlining Gingival and Mucosal Tissues}
For ROI selection, the largest possible ROIs were selected for each tissue, excluding regions of hard tissues such as bone, crown as well as the epithelial layer and rete pegs, as those will affect the QUS parameter estimations representing gingiva and mucosa characteristics The ROI selection for the QUS analysis in each scan was checked with specialists and periodontists to make sure ROIs are selected properly for each tissue type. Distinction of the two tissue types adjacent to each other during the ROI selection was based on two factors: spatial location of landmarks such as bone and epithelium as well as visual speckle pattern differences of gingiva and alveolar mucosa. It is also noted that a small spatial gap was assumed between ROIs in gingiva and alveolar mucosa during ROI placement to prevent including any possible overlapping region.

\subsection{\textbf{Histology Images}}
After acquiring \textit{in vivo} intraoral US scan data, animals were euthanized and tissue block samples were collected from US scanning sites. Samples incorporated oral soft and hard tissues such as alveolar mucosa, gingiva, bone and crown, allowing for transverse cross-sectional cuts of oral sites to be a part of histology images as also observed via US imaging. Tissue samples were collected and immersed in 10\% formalin to be preserved from decay and maintain tissue structure for further tissue staining and processing. Tissue blocks were placed in 10\% EDTA (Ethylenediaminetetraacetic acid) for demineralization for 3 to 6 months and then embedded within paraffin wax to maintain its shape as well as to create a support frame and facilitate tissue slicing (5 micron-thickness). Slices were stained using two separate methods: Masson's Thrichrom and H\&E techniques. \newline
The Masson's Trichrome stain is a three-color staining method employed to reveal collagen structures of hard and soft tissues with its signature blue color for collagen. In this stain, red represents cytoplasm/red blood cell and dark purple/black shows nuclei. For example, dentine with its dominant collagen matrix is stained blue and keratin is stained as red. On the other hand, H\&E as the most common staining method, is a two-color stain method that displays the underlying tissue morphology with a purplish color for nuclei and varying shades of pink color for cytoplasm, extracellular matrix and other structures.
Stained tissue slices were imaged using an optical microscope (E800, Nikon Instruments Inc., Melville, NY) with 4x and 20x magnifications for overall and local imaging of slices, respectively.
%%%%%%%%%% Results
\section{Results}
\label{Results}
\subsection{\textbf{Phantom Study}}

\begin{figure}[b]
\centering
\includegraphics[width=0.87\textwidth, trim={0.9in 2.2in 0.6in 4in}] {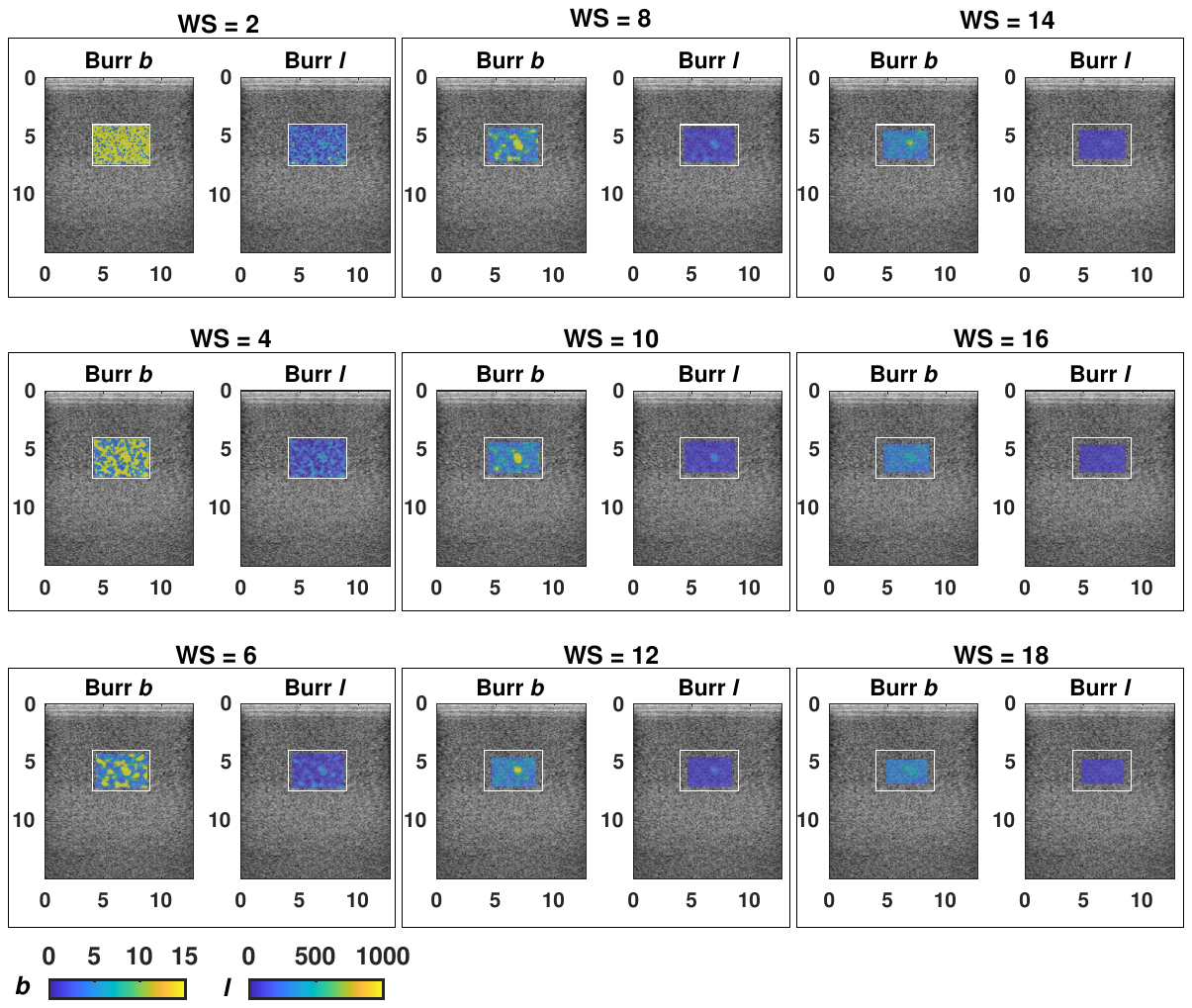}% change 'example-image' to the filename of your image
\caption{Burr parametric imaging for varying window size (WS in multiples of wavelength) superimposed on the associated B-mode phantom image. In each pair of parametric images, the left image shows Burr $b$ and the right one shows Burr $l$. All axes are shown in millimeters.}
\label{cirs burr}
\end{figure}

In parametric images, linear interpolations were performed between local estimations of adjacent centers to obtain smoother presentations. This interpolation did not result in any significant variation in statistical parameter estimations. For a selected ROI comparing linear interpolation vs. no interpolation, \textit{p-values} were assessed using the non-parametric Wilcoxon Rank Sum Test (a.k.a. Mann-Whitney U test), as the data were not normally distributed. Results showed that for the Burr \textit{b} and \textit{l}, \textit{p-values} were 0.29 and 0.28, respectively and for the Nakagami \textit{m} and $\Omega$, \textit{p-values} were estimated as 0.82 and 0.57. These results (\textit{p-values} $>$ 0.05) indicate that the linear interpolation does not impose any statistical significance in the estimated model parameters.  \newline
\textbf{Figure \ref{cirs burr}} shows nine pairs of parametric images for the Burr two parameters superimposed onto the same B-scan image for nine WSs. In each pair of parametric images, the left-side image represents the Burr power-law parameter $b$ and the right-side one shows the Burr scale factor $l$. Burr parameters were estimated within a rectangular (5 mm by 2.5 mm) ROI shown as white solid box. The dynamic range of all colorbars were set equal to provide for an absolute comparison of results across all WSs. \newline
It is observed that the sliding WS of 2 produces parametric images with highly fluctuating statistical estimations, showcasing that local estimations may vary widely over a large dynamic range (color saturations beyond the colorbar range are observed). This is consistent with the expectation that speckle data from small kernels highly fluctuates, i.e., statistical estimations from excessively small kernels are not robust. By increasing WS from 2 to 10 wavelengths, we observe an improvement in homogeneity of both Burr parameters, $b$ and $l$. Comparing parametric images produced from WS of 10 to 16 show the presence of a small non-homogeneous region (visible as a yellow patch) where local Burr $b$ and $l$ are persistently higher. Therefore, these local estimations are independent of the size of sliding kernel and are associated with local variations in underlying phantom structures from the Burr model standpoint. It is noted that for the sliding kernel size of 16 and 18, it transitions from a local parameter estimator to a global estimator over the whole ROI. This global estimation manifests itself as a parametric image being too uniform in which information on local variations are lost (averaged out).

\begin{figure}[h]
\centering
\includegraphics[width=0.87\textwidth, trim={0.9in 2.2in 0.6in 4in}] {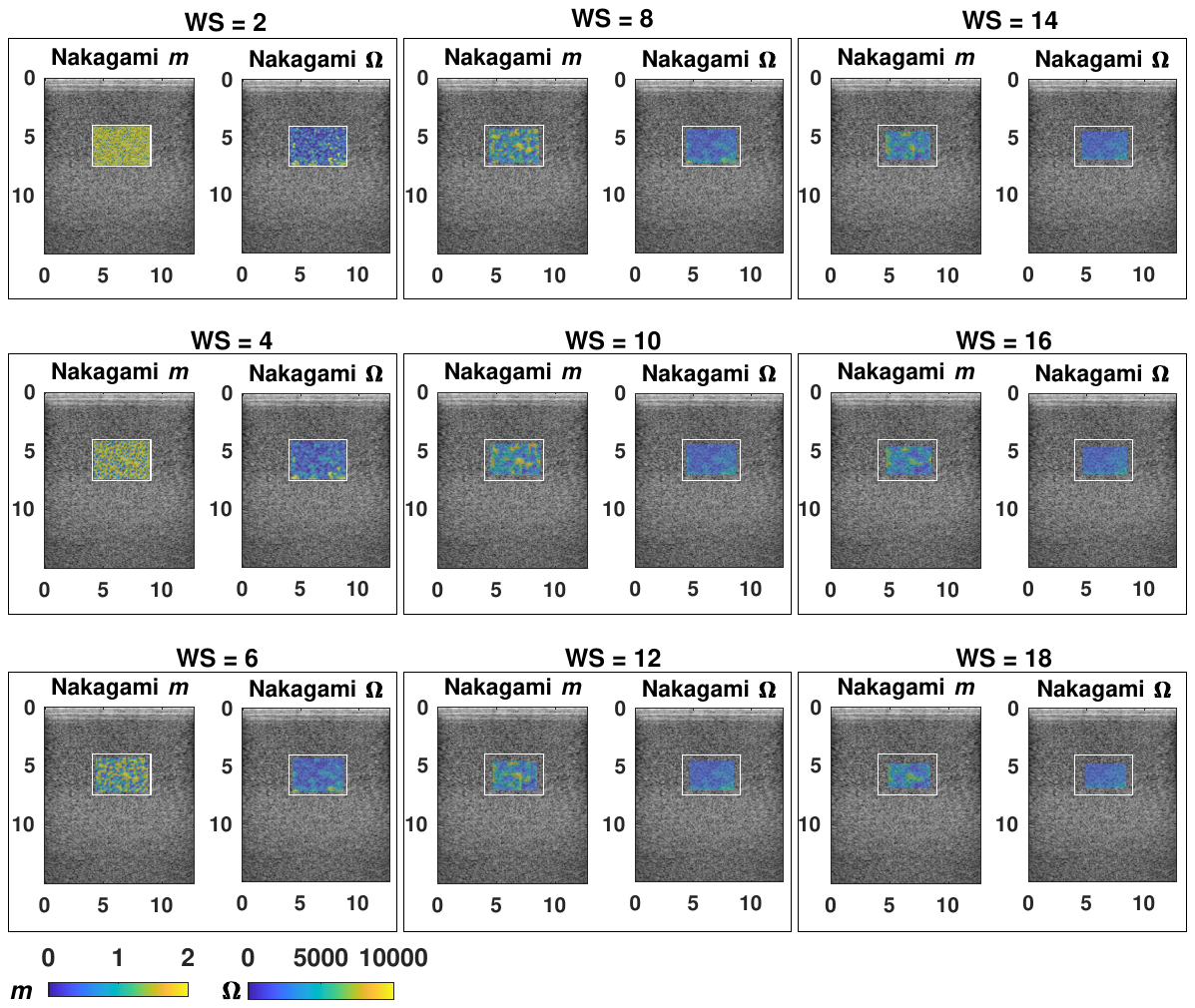}% change 'example-image' to the filename of your image
\caption{Nakagami parametric imaging for varying window size (same phantom study as shown in \textbf{Figure \ref{cirs burr}}). In each
pair, the left image shows Nakagami $m$ and the right one shows Nakagami $\Omega$ parameter local estimations. All axes are shown in millimeters.}
\label{cirs naka}
\end{figure}

\begin{figure}[h]
\centering
\includegraphics[width=1\textwidth, trim={0.7in 2.5in 1in 4in}] {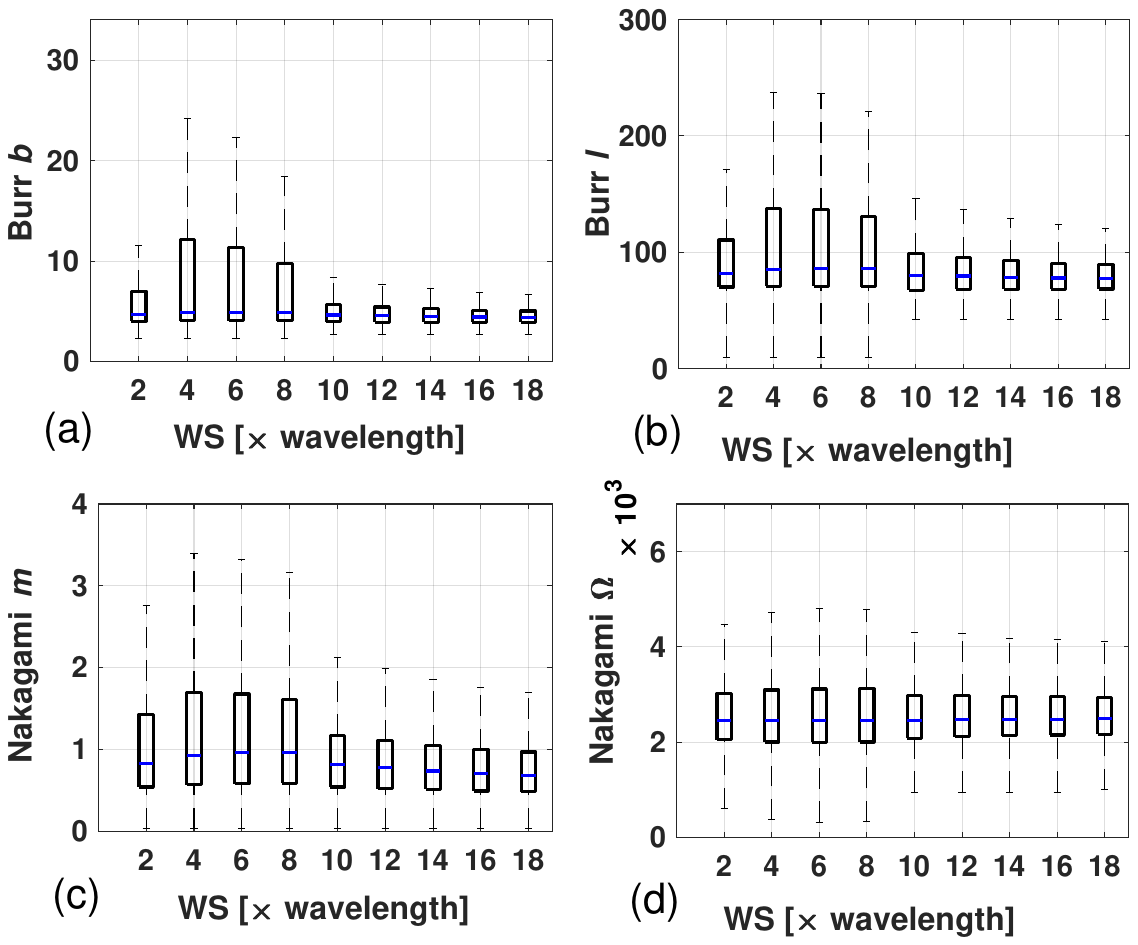}% change 'example-image' to the filename of your image
\caption{Boxplots showing the effect of WS on the Burr (top row), and the Nakagami (bottom row) parameters. Results in boxplots are from four different imaging planes within an ROI size of 5 mm by 2.5 mm (shown in \textbf{Figure \ref{cirs burr}} and \textbf{Figure \ref{cirs naka}}).}
\label{ws}
\end{figure}

\textbf{Figure \ref{cirs naka}} shows parametric images for the Nakagami shape parameter \textit{m} and scale factor $\Omega$ for different WSs, similar to what was done with the Burr model in \textbf{Figure \ref{cirs burr}}. For the Nakagami model, parametric images show that at a WS of 10 wavelengths and more, variations in local parameter estimations are not as intense as for smaller kernel sizes. Similar to the Burr model, there exists a small non-uniformity (shown as color saturation beyond the colorbar range) in parametric images, independent of the WS. A more detailed investigation into the effect of WS on the Burr and Nakagami parameter estimation is presented in \textbf{Figure \ref{ws}} as boxplots. At each WS, the boxplot represents estimations from four different scans of the phantom. In this figure, \textbf{(a)} and \textbf{(b)} show results for the Burr \textit{b} and \textit{l}, respectively. \textbf{(c)} and \textbf{(d)} are estimations for the Nakagami \textit{m} and $\Omega$ in that order. Results show that variations (such as interquartile range) in Burr \textit{b} and \textit{l} as well as Nakagami \textit{m} are larger at smaller WSs. At WS$=$10 wavelengths (wavelength is 64.2 µm at 24 MHz) and higher, these variations become smaller (this effect is less pronounced for the Nakagami $\Omega$, which essentially represents echo intensity). For instance, comparing percent differences ($ |X-Y|/(X+Y)/2) \times 100) $ for Burr \textit{b}, the upper quartile at WS=10 decreases approximately by 52$\%$ compared to WS=8 where it shows a variation of less than 6\% when even comparing to WS=12. Model parameters converge, and their variations become small with further increase in WS$=$10. Specifically, for WSs larger than 10 wavelengths,  their median (and the upper and lower quartiles) values vary less than 3.5$\%$ for the two scale factors and less than 6$\%$ for the Burr power-law and Nakagami shape parameters. Thus, WS=10 provides sufficient speckle data for a reasonable and robust statistical parameter estimation. This confirms visual assessments of the WS effect on parametric images.
As a note on the optimal WS selection from a practical standpoint where we deal with relatively small-sized periodontal soft tissues, the WS of 10 wavelengths reasonably meets criteria of tissue size limitation and precision.  Therefore, it is suggested as the optimal kernel size for the Burr and Nakagami parametric imaging of swine tissues.
\\
\subsection{\textbf{Parametric Imaging for \textit{In vivo} Swine Scans}}
\begin{figure}[h]
\centering
\includegraphics[width=0.9\textwidth, trim={1in 3.3in 1in 0.2in}] {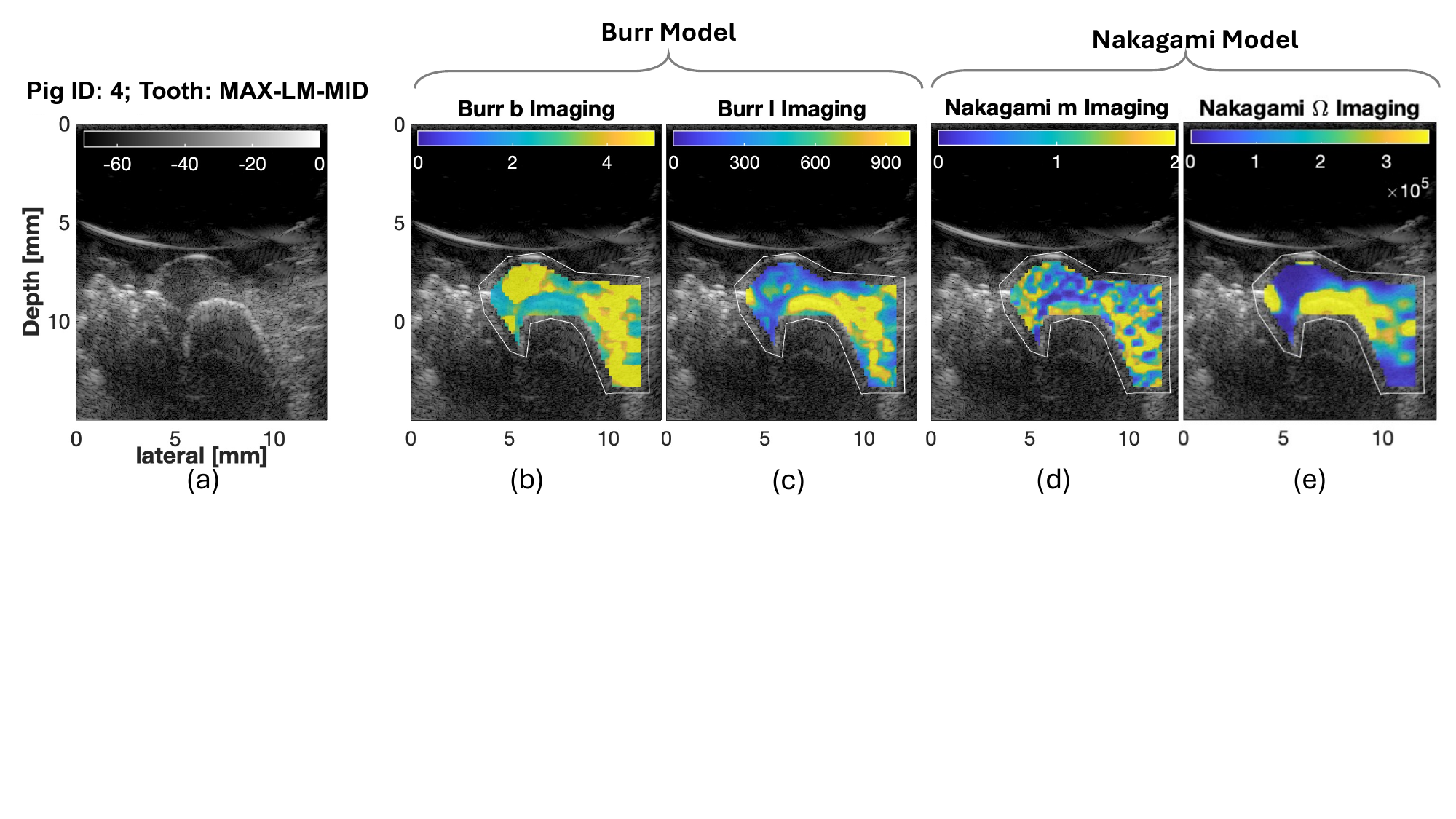}% change 'example-image' to the filename of your image
\caption{Parametric imaging of the Burr and Nakagami parameters in periodontal soft
tissues in a swine scan using the WS of 10 wavelengths, for maxilla left first molar tooth.
\textbf{(a)} reference B-scan, \textbf{(b)} Burr power-law parameter $b$ and \textbf{(c)} Burr scale factor $l$, \textbf{(d)} Nakagami shape parameter $m$ and \textbf{(e)} Nakagami scale factor $\Omega$.}
\label{PI}
\end{figure}

\begin{figure}[h]
\centering
\includegraphics[width=0.9\textwidth, trim={1in 0.3in 2.8in 0.5in}] {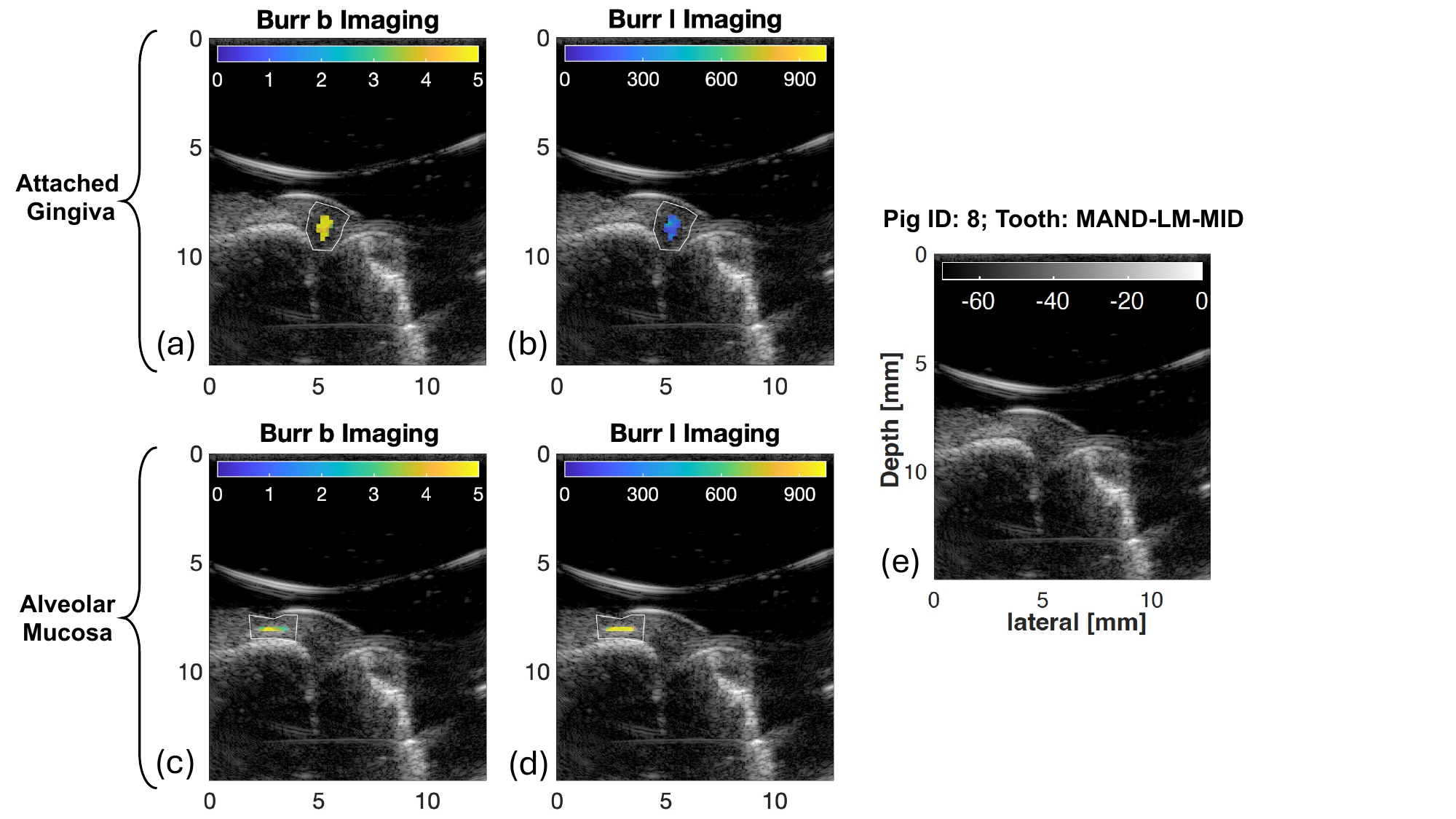}% change 'example-image' to the filename of your image
\caption{Parametric imaging of the Burr parameters in the gingiva (top row) and the alveolar mucosa (bottom row) in a swine scan using the WS of 10 wavelengths. \textbf{(a)} and \textbf{(c)}: Burr $b$, \textbf{(b)} and \textbf{(d)}: Burr $l$. The reference B-scan in shown in \textbf{(e)}.}
\label{PI Burr}
\end{figure}

\begin{figure}[h]
\centering
\includegraphics[width=0.9\textwidth, trim={1in 0.3in 2.8in 0.5in}] {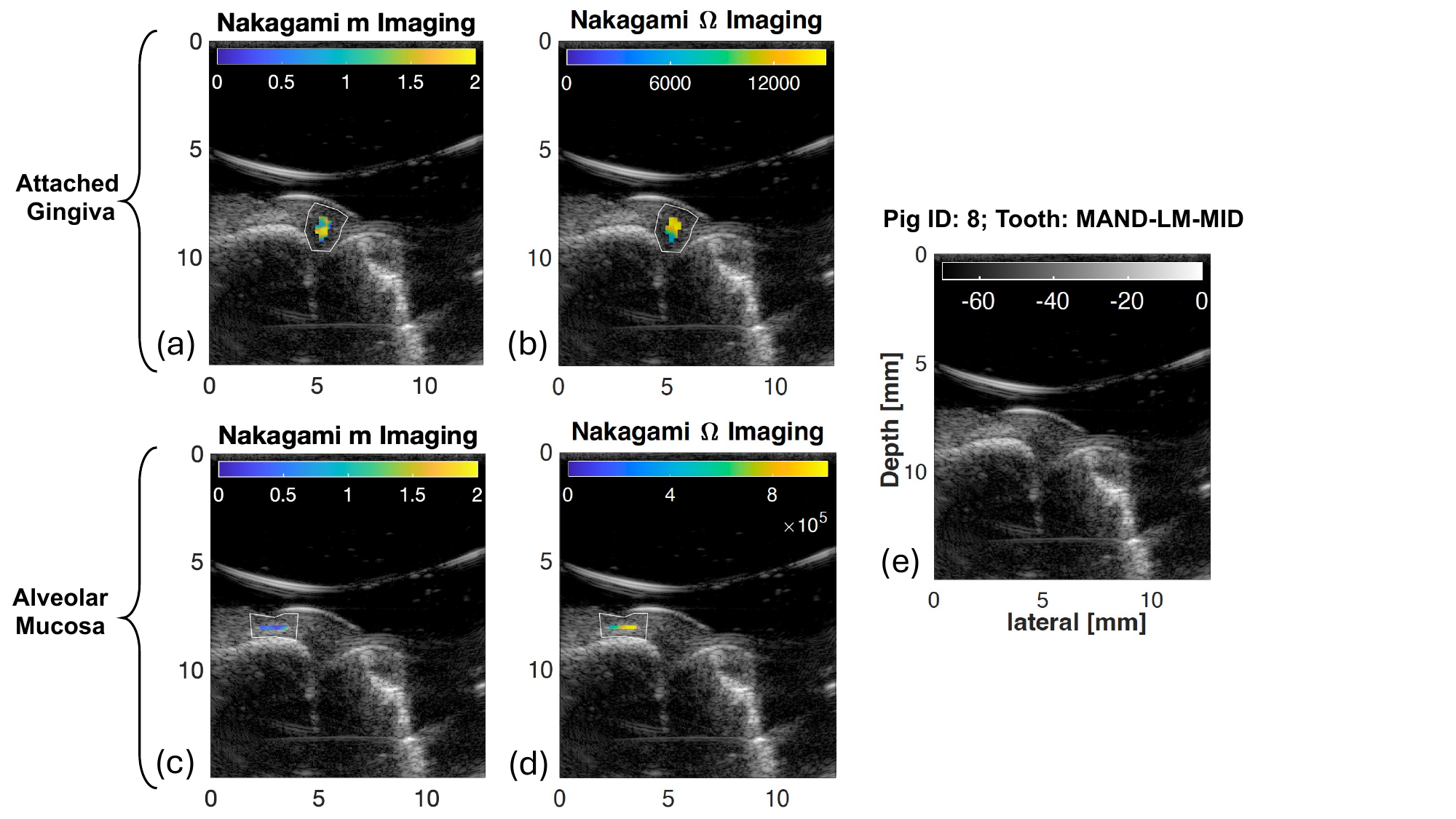}% change 'example-image' to the filename of your image
\caption{Parametric imaging of the Nakagami parameters in the gingiva (top row) and the alveolar mucosa (bottom row) in a swine scan using the WS of 10 wavelengths. \textbf{(a)} and \textbf{(c)}: Nakagami $m$, \textbf{(b)} and \textbf{(d)}: Nakagami $\Omega$. The reference B-scan in shown in \textbf{(e)}.}
\label{PI Naka}
\end{figure}

The parametric images of the Burr and Nakagami parameters of periodontal soft tissues were obtained in ultrasound scans of 10 swine at the baseline condition (no inflammation) using the optimal sliding WS of 10 wavelengths. The results were statistically compared in gingiva versus alveolar mucosa within reasonable ROIs selected for each scan. An example of parametric image of the Burr $b$, Burr $l$, Nakagami $m$ and Nakagami $\Omega$ overlaid on corresponding B-scans are shown in \textbf{Figure \ref{PI}} with the reference B-scan. The ROIs selected for these parametric images include the marginal and attached gingiva, alveolar mucosa, epithelium as well as muscle tissues and is meant to show the variations of estimated parameters over the whole scanned region. However, for the quantitative characterization of gingiva and alveolar mucosa, individual ROIs are selected for each tissue excluding any bone, epithelium layers or other regions not associated with these tissues. 
An example of the ROI selection from these two tissue types is shown in \textbf{Figure \ref{PI Burr}} and \textbf{Figure \ref{PI Naka}} for the Burr and Nakagami parametric imaging. In these two figures, top rows represent ROI for the gingiva and bottom rows show alveolar mucosal ROI, with the reference B-scan is presented in \textbf{(e)}. It is observed that for the gingival tissue, the estimated Burr power-law parameter $b$ and the Nakagami shape parameter $m$ map to a brighter (yellow) colormap, corresponding to higher estimated values compared to mucosal tissues. On the other hand, the Burr scale factor $l$ and the Nakagami scale factor $\Omega$ for gingival tissues are lower than mucosal tissues based on the colormap comparison.

\subsection{\textbf{Statistical Analysis of Swine Populations}}
\begin{figure}[h]
\centering
\includegraphics[width=0.3\textwidth, trim={3.5in 2.4in 3in 3in}] {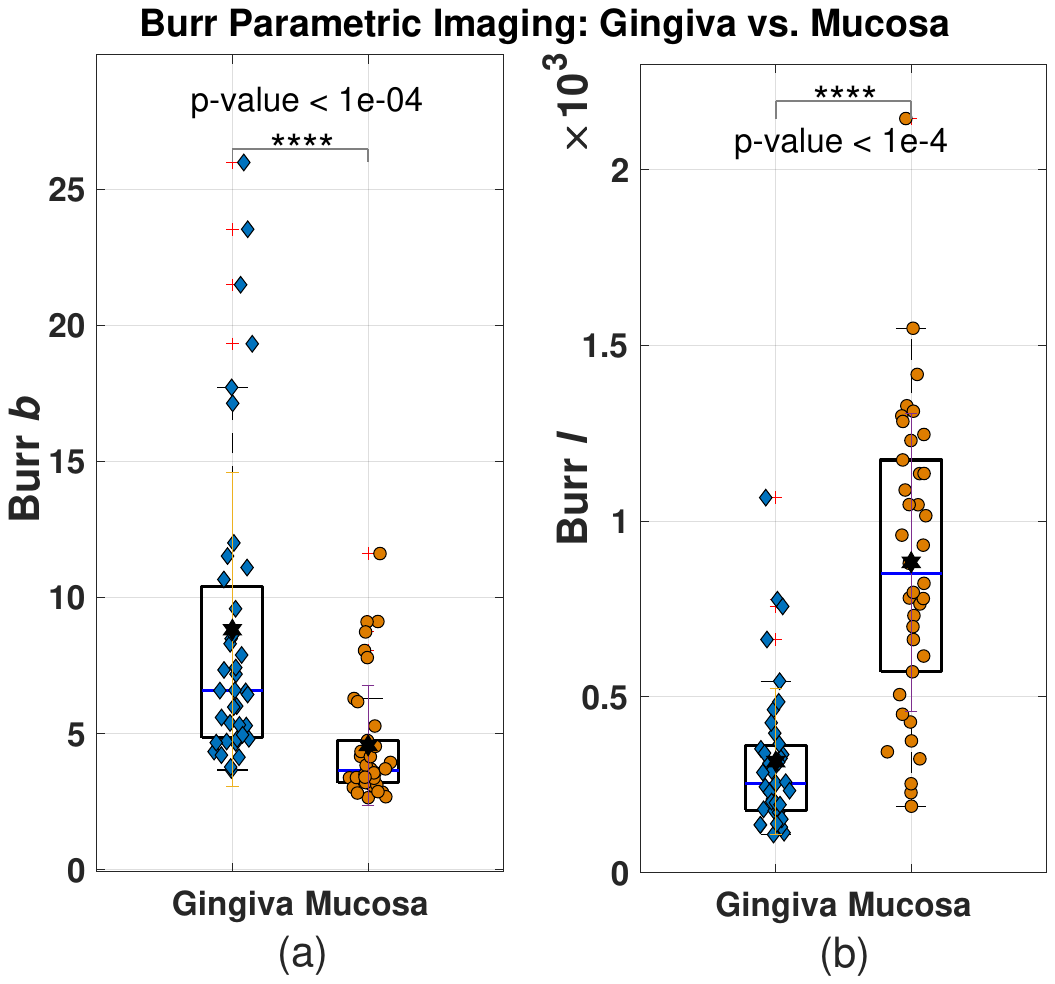}% change 'example-image' to the filename of your image
\caption{Burr parameters for classification of gingiva vs. alveolar mucosa, with boxplots summarizing the average estimations within ROIs in swine cases. \textbf{(a)} Burr $b$, \textbf{(b)} Burr $l$. In each boxplot, the blue line shows the statistical median for the population and the pentagon symbols represent the statistical mean of the  estimations.}
\label{Burr boxplot}
\end{figure}

\begin{figure}[h]
\centering
\includegraphics[width=0.3\textwidth, trim={3.5in 2.4in 3in 3in}] {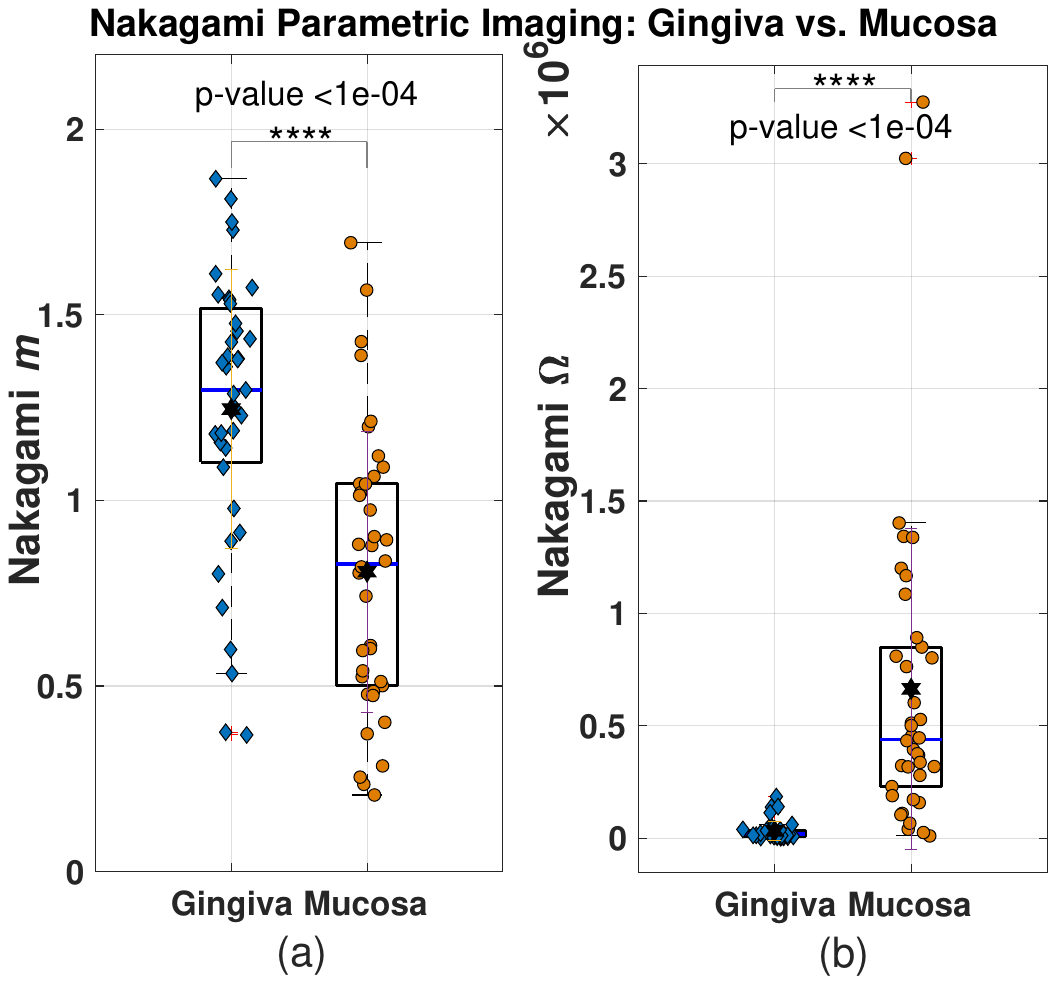}% change 'example-image' to the filename of your image
\caption{Nakagami parameters for classification of gingiva vs. alveolar mucosa, with boxplots summarizing the average estimations within ROIs in swine cases. \textbf{(a)} Nakagami $m$, \textbf{(b)} Nakagami $\Omega$. In each boxplot, the blue line shows the statistical median for the population and the pentagon symbols represent the statistical mean of the  estimations.}
\label{Naka boxplot}
\end{figure}

To characterize these two types of periodontal soft tissues adjacent to each other at the baseline condition (no inflammation) for all swine populations, the summary of QUS analyses are presented as boxplots for the Burr and Nakagami models in \textbf{Figure \ref{Burr boxplot}} and \textbf{Figure \ref{Naka boxplot}}, respectively. In these figures, $p-values$ are reported to compare statistical significance between the two tissue types. In each boxplot, the blue line shows the statistical median for the population and the pentagon symbols represent the statistical mean of the  estimations. \newline
In \textbf{Figure \ref{Burr boxplot}}, \textbf{(a)} and \textbf{(b)} shows the Burr $b$ and Burr $l$ results. For the Burr parameters, $p-values<0.0001$, indicating that $b$ and $l$ could show statistically significant populations when comparing gingival tissues with alveolar mucosa. The Burr $b$ for gingiva is reported to be higher than mucosa when comparing mode (Quartile1 $|$ Quartile3) of results in gingiva vs. alveolar mucosa  ($b_{Gingiva}=6.6\:(4.8|10.4)$, $b_{Mucosa}=3.6\:(3.2|4.7)$). On the other hand, the Burr $l$ is reported as being lower in gingival tissues than in the muscosa ($l_{Gingiva}=254.0\:(177.1|362.7)$, $l_{Mucosa}=851.8\:(571.6|1174.3))$. In \textbf{Figure \ref{Naka boxplot}} for the Nakagami model, \textbf{(a)} shows that Nakagami $m$ is statistically higher for gingiva population in comparison to the alveolar mucosa ($m_{Gingiva}=1.29\:(1.10|1.52)$, $m_{Mucosa}=0.83\:(0.50|1.04))$. \textbf{(b)} represents a statistically elevated Nakagami $\Omega$ in alveolar mucosa compared to the gingiva ($\Omega_{Gingiva}=1.82\times 10^4\:(0.81\times 10^4|3.57\times 10^4)$, $\Omega_{Mucosa}=4.39\times 10^5\:(2.29\times 10^5|8.50\times 10^5))$. Nakagami $\Omega$ clearly demonstrates that echo intensity in gingiva is significantly lower (an order of magnitude).

\subsection{\textbf{Histology Insight}}
To cast some insight into a possible explanation for statistically distinct QUS parameters for alveolar mucosa and gingiva, we consulted their histology images. The results for Masson’s Trichrome and H\&E stains are shown in \textbf{Figure \ref{Trichrom}} and \textbf{Figure \ref{H&E}}, respectively. In these figures, \textbf{(a)} represents histology insights (4x magnification) of the stained tissue slice incorporating both soft and hard tissues. \textbf{(b)} and \textbf{(c)} show a 20x magnified image of gingival and alveolar mucosal tissues, respectively, marked with red dashed box in \textbf{(a)}. Scale bars are also added to all images for references.
\begin{figure}[h]
\centering
\includegraphics[width=0.95\textwidth, trim={0.3in 0.3in 0.3in 0.3in}] {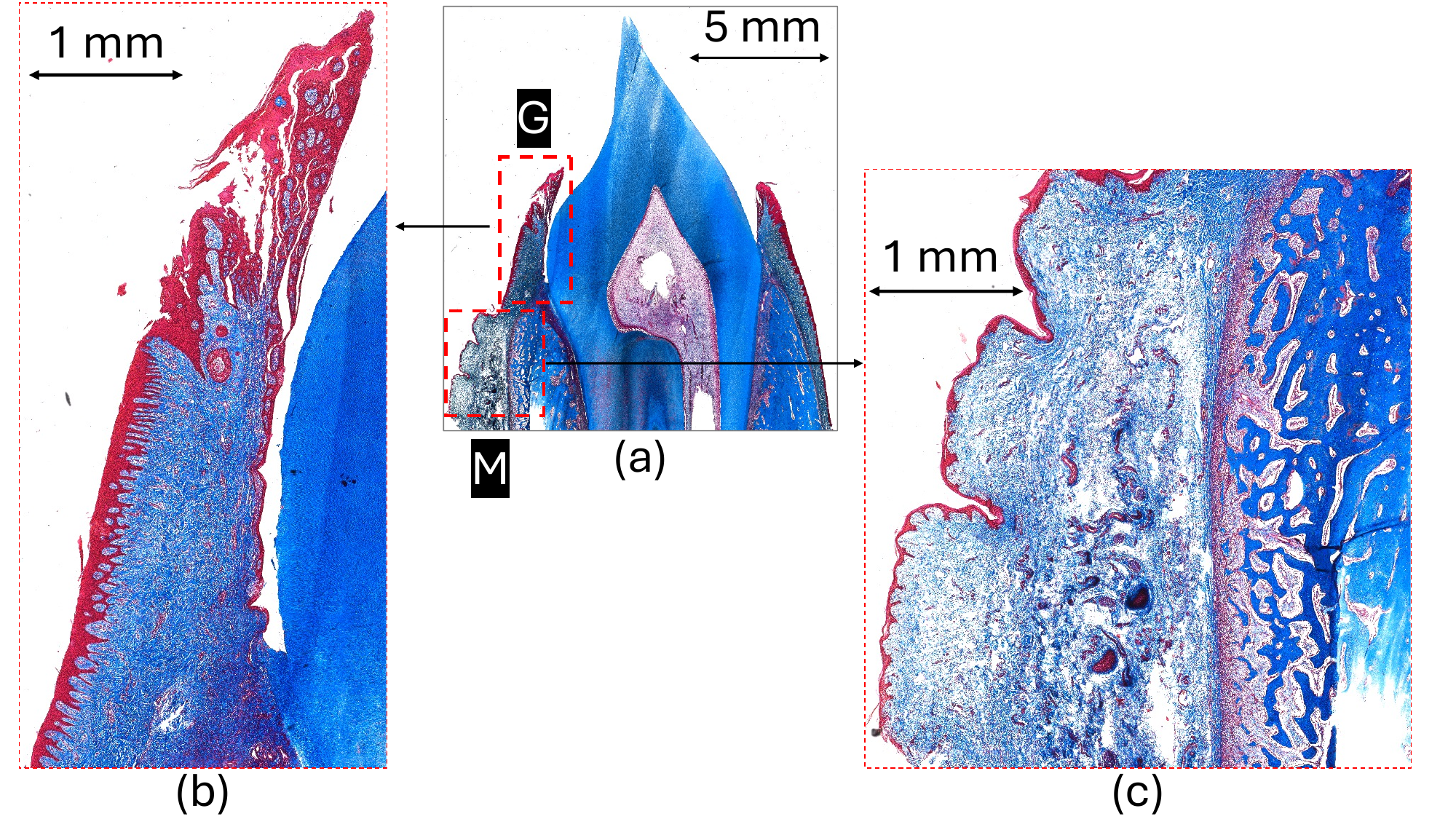}% change 'example-image' to the filename of your image % Daria: Histology will be added
\caption{\textbf{(a)} Histology image using Masson’s Trichrome stain (4x magnification). \textbf{(b)} and \textbf{(c)} are enlarged views comparing gingival and alveolar mucosal regions, respectively (20x magnification). Swine oral site: left mandibular.}
\label{Trichrom}
\end{figure}
\begin{figure}[h]
\centering
\includegraphics[width=0.95\textwidth, trim={0.3in 1in 0.3in 0.1in}] {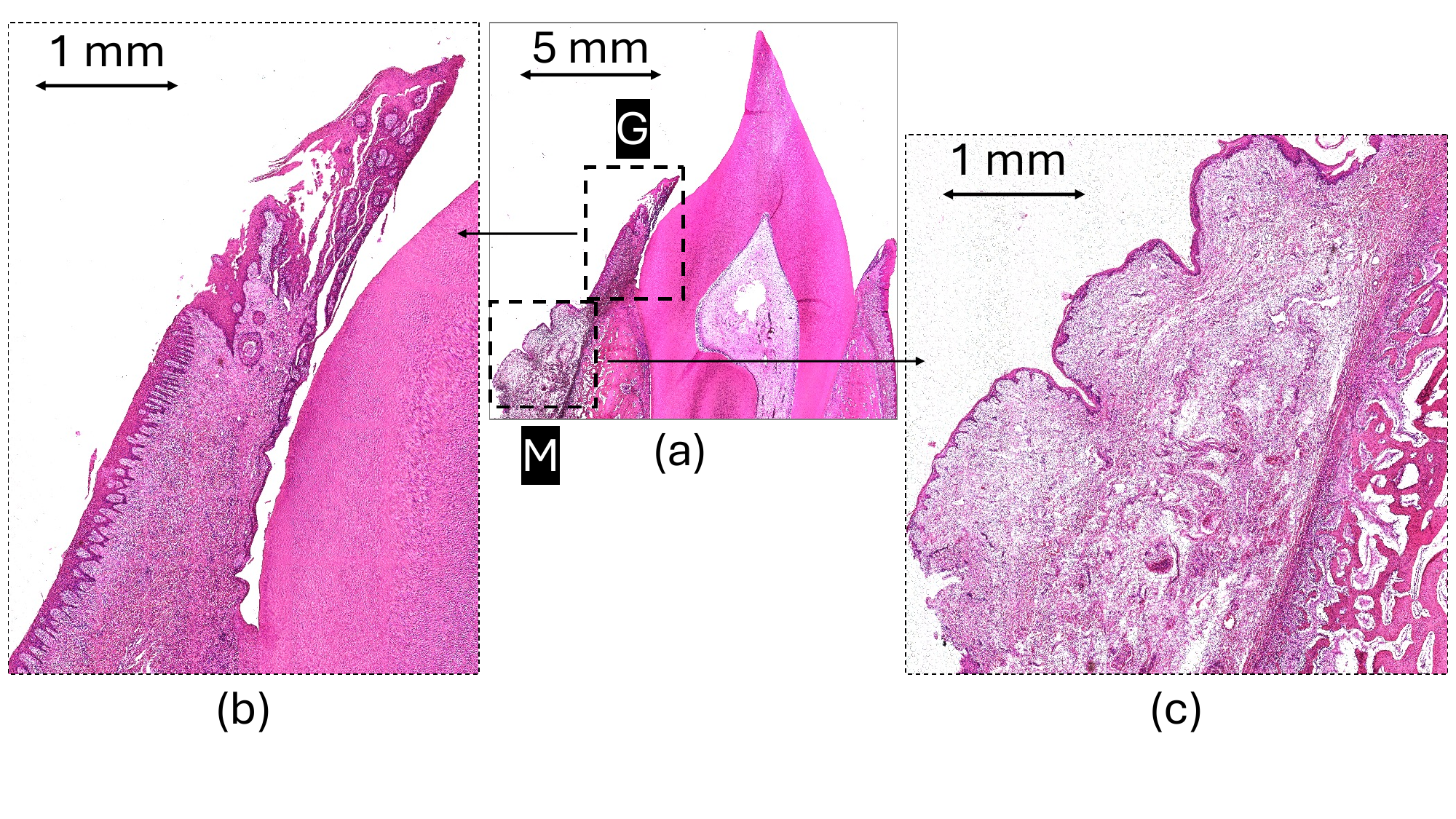}% change 'example-image' to the filename of your image % Daria: Histology will be added
\caption{\textbf{(a)} Histology image using H\&E stain (4x magnification). \textbf{(b)} and \textbf{(c)} are enlarged views comparing gingival and alveolar mucosal regions, respectively (20x magnification). Swine oral site: left mandibular.}
\label{Trichrom}
\end{figure}
\noindent Looking at histology images from both staining techniques, we notice a denser stain in gingival regions compared to alveolar mucosa, which could suggest denser scattering sites in gingiva.

\subsection{\textbf{QUS-based classifications of alveolar mucosa and gingiva}}
\begin{figure}
\centering
\includegraphics [width=1\textwidth, trim={1.3in 0.2in 1.3in 2.2in}] {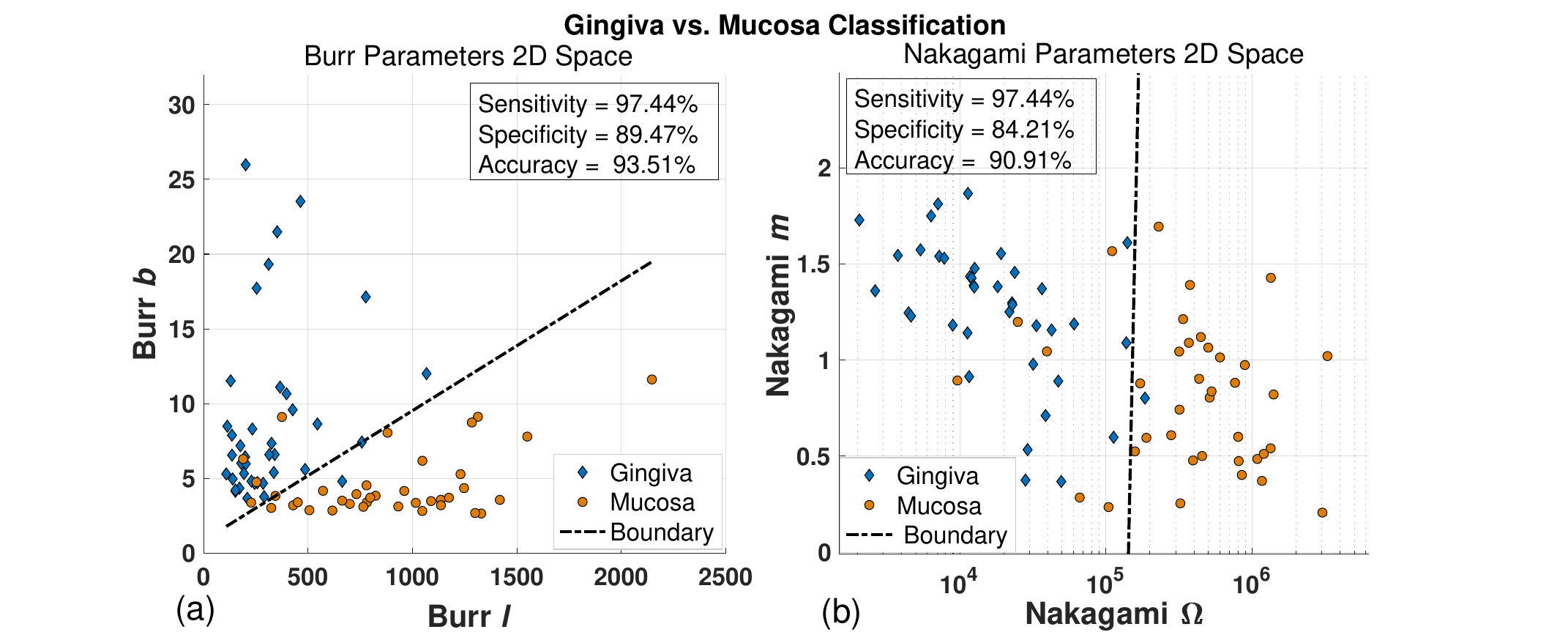} % Daria: problem with importing tif.
\caption{2D classifications of gingiva vs. alveolar mucosa in swine cases using \textbf{(a)} the Burr model (Burr $b$ vs. Burr $l$), and \textbf{(b)} the Nakagami model (Nakagami $m$ vs. Nakagami $\Omega$). Black lines show linear boundaries between two classes of tissues. Blue diamond symbols: estimations for gingiva, orange circle symbols: estimations for mucosa.}
\label{2D Class}
\end{figure}

\begin{figure}[ht]
\centering
\includegraphics[width=0.4\textwidth, trim={1.5in 2.8in 3.2in 2.5in}] {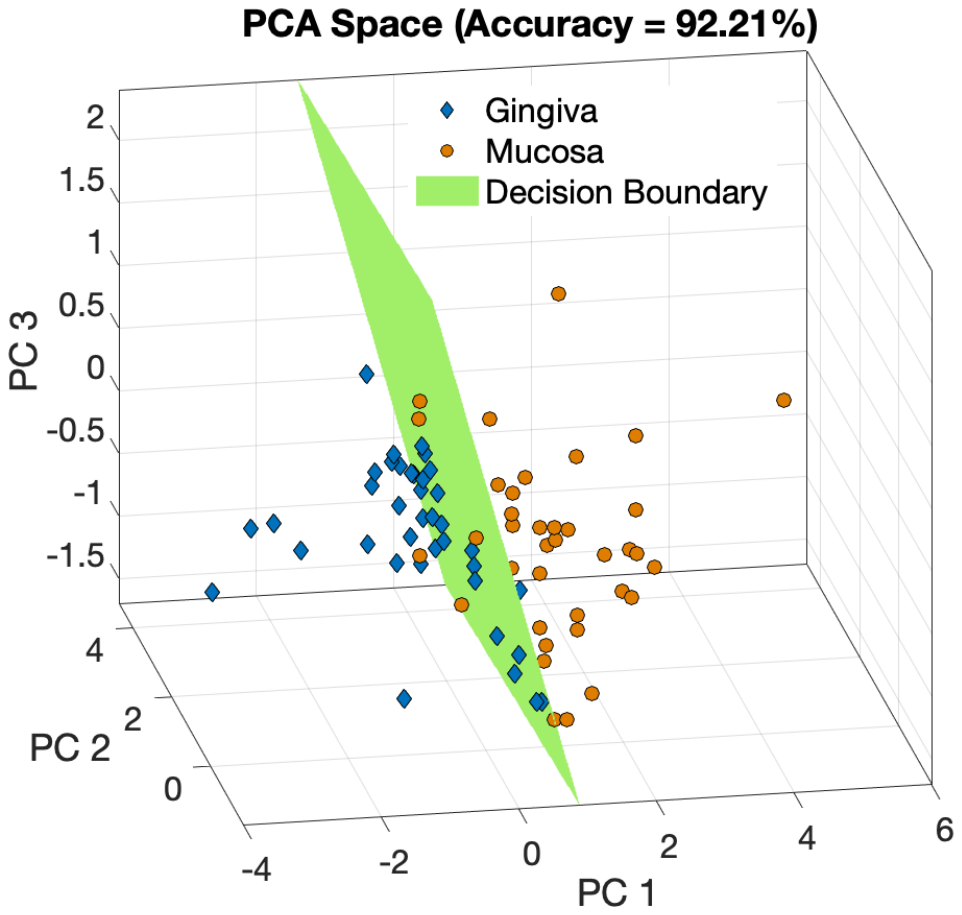}% change 'example-image' to the filename of your image
\caption{3D Classification of gingiva vs. alveolar mucosa using combined parameters of the Burr and Nakagami models, represented in the principal component space using the first three principal components of PC 1, PC 2, and PC 3.}
\label{3D Class}
\end{figure}

To further investigate the separations of alveolar mucosal and gingival tissues from the QUS standpoint, averages of the Burr and Nakagami parameters are estimated for each swine from local statistical estimations. The results are illustrated as two separate 2D scatterer plots of $l-b$ and $m-\Omega$ in \textbf{Figure \ref{2D Class}} \textbf{(a)} and \textbf{(b)}, respectively. In \textbf{(b)}, the Nakgami $\Omega$ axis is represented as log-scale to compress the large dynamic ranges of this parameter in alveolar mucosa and gingiva.
In these figures, orange circles represent the estimations for alveolar mucosa and blue diamond symbols show the estimations for gingiva. In each 2D space, a linear boundary is optimized to classify the 2D parameter space by maximizing the accuracy of tissue type prediction. These lines show a clear separation of mucosal and gingival tissues in 2D space of QUS parameters estimated from the US speckle statistics. For the Burr model, the linear classification line gives an accuracy of 93.51\%, a sensitivity (true positive rate) of 97.44\%, and a specificity (true negative rate) of 89.47\% (gingival tissue is assumed to be the positive class). For the Nakagami model, accuracy, sensitivity and specificity are 90.91\%, 97.44\%, and 84.21\%, respectively.
Further, the classification of the two tissue types is investigated by combining the Burr and Nakagami parameters resulting in four features of $b$, $l$, $m$, and $\Omega$, to assess the multi-parametric classification accuracy. To represent the clustering of the two classes with four parameters in a reduced dimensionality (3D) space, the principal component analysis (PCA) is performed on the parameter (feature) space to map the original parameters into a new set of variables, a.k.a. principal components (PCs). PCs are basically directions in the feature space, composed of a linear combination of the original features, along which the data shows the most variation (higher variance). The maximum number of PCs in this case is four, however, we employ the first three PCs to visualize data in 3D. These PCs are the most significant representations of variations (dispersion) in the data. The mapped data onto the PC space along with the mapped classification boundaries are shown in \textbf{Figure \ref{3D Class}}. The variance of data captured by the PC 1 is 55.31\%, which is the direction with the highest dispersion in the data. PC 2 and PC 3 explain 25.37\% and 12.38\% of the variance in the data, respectively. Therefore, the first three PCs used to visualize the data in 3D PC space represent 93.06\% of the variance of the data. The data are standardized before being mapped to the PC space to be scaled. Therefore, the axes range in this figure are different than the range of original parameters. The decision boundary using the four parameters gives the classification accuracy of 92.21\%, which is close to the two classification accuracies when applying each model separately.
\textbf{}
%%%%%%%%%%% DISCUSSION
\section{Discussion}
\label{Discuss}
Our phantom study suggested that WS=10 wavelengths was a reasonable kernel size for local statistical estimation of the Burr and Nakagami parameters, particularly for small-sized periodontal tissues. It is noted that the higher range of variations consistently observed even for the largest WS may be related to the underlying structure of the phantom and might not represent a WS dependence per se. Our findings from QUS analysis of gingiva and alveolar mucosal tissues suggest that Burr and Nakagami parameters show sensitivities to these periodontal tissue types, as supported by \textbf{Figure \ref{PI Burr}}, and \textbf{Figure \ref{PI Naka}} for parametric images and \textbf{Figure \ref{Burr boxplot}} and \textbf{Figure \ref{Naka boxplot}} for the swine population. Thus, QUS hold potentials to characterize periodontal tissues. It is noted that the range of the Burr $b$,  which characterize the power-law nature of the scattering structure (scatterer number density), is reported to be close to 3 for some soft tissues such as liver \cite{RN4, RN5}. In this study, the median for the alveolar mucosal tissue tends to be close to this range while this is elevated for gingival tissues.\newline
In this study, our preliminary investigation into multi-parametric characterization of the two tissue types using combined Burr and Nakagami parameters provided a high classification accuracy but close to the 2D classification accuracies. The multi-parametric analysis accuracy did not significantly improve compared with the 2D classifications. The main potential reason behind this is the fact that multi-parametric classification is done based on maximizing the boundary margins whereas the 2D classification is performed by focusing on maximizing the accuracy itself directly rather than separation margins. Another potential explanation is the possible overlap in information explained by the two models. For both model, linear classification lines accurately classify all but one gingival case, leading to obtaining similar estimations for sensitivities, however the Burr model estimates more accurate alveolar mucosal cases, resulting in higher specificity than the Nakagami model. By comparing these statistics, the Burr model represents a relatively better separation of the two tissue types, with a slightly higher accuracy and also specificity. It is also noted that PCs are less interpretable representation of the data, and they mostly serve as an effective mean to represent the data in lower dimensional space while retaining most essential information from the data. For a classification with higher number of features, the PCA makes training of the classification model more effective. \newline
We suggest that QUS findings for periodontal tissue characterization are supported by qualitative analysis of histology images. As introduced in the Introduction, various fiber types exist in the gingiva and alveolar mucosa. They also present in different arrangements with their primary role of protecting the structure of root, bone and soft tissues from degradation and deformation. In ultrasound imaging of periodontal tissues, potential sources of acoustic scattering affecting image formation include fibers as well as vessel walls. Comparing the two tissue types in histology images with statistical QUS findings, observations suggest some consistencies between these findings. Results might align in two aspects, as outlined in following paragraphs.
\begin{itemize}
    \item First, lower B-scan echo intensity in gingiva compared to alveolar mucosa (indicated by decreased Burr and Nakagami scale factors) is hypothesized to arise from occurrence of multiple scattering between densely-packed scattering sites in gingiva. This could result in a weaker backscatter signal from this region back to the transducer. Thus, gingiva would appear less echogenic in US B-scans compared to alveolar mucosa.
   \item Second, the histology finding may suggest the presence of higher densities of small, densely-packed scatterers. This could explain elevated estimations of Burr power-law parameter $b$ (associated with the number density of scatterers) obtained for the gingiva compared to the alveolar mucosa in the QUS analysis. Additionally, the QUS analysis showed a higher Nakagami shape parameter $m$ in gingiva compared to alveolar mucosa, indicating a transition of scattering statistics to Rayleigh and post-Rayleigh regime for gingiva. The Rayleigh scattering regime is characterized by the presence of many random small scatterering sites, resulting in a diffuse scattering. Thus, the elevated Nakagami shape parameter in gingiva could be associated with the denser stain observed in the gingiva on histology images, corresponding to an increase in scattering number density, (transitions into the Rayleigh and post-Rayleigh regime).
   \indent The PDF of the US speckle data within ROIs from alveolar mucosa and gingiva are fitted to the Rayleigh distribution in \textbf{Figure \ref{PDF}} as an example of PDF comparison. It is observed that speckle statistics of the gingiva fits relatively better to the Rayleigh ($R-squared=0.95$) compared the alveolar mucosa ($R-squared=0.82$). This suggests a gradual transition towards Rayleigh model for gingiva compared to alveolar mucosa. In these PDFs, the Burr model is added as a reference speckle model and it is noteworthy that it is a more accurate fit to statistics of both tissue types with its distinctive heavy-tail behavior at the higher amplitudes compared to the Rayleigh model.
\begin{figure}[ht]
\centering
\includegraphics[width=1\textwidth, trim={0.7in 1.8in 3.2in 8in}] {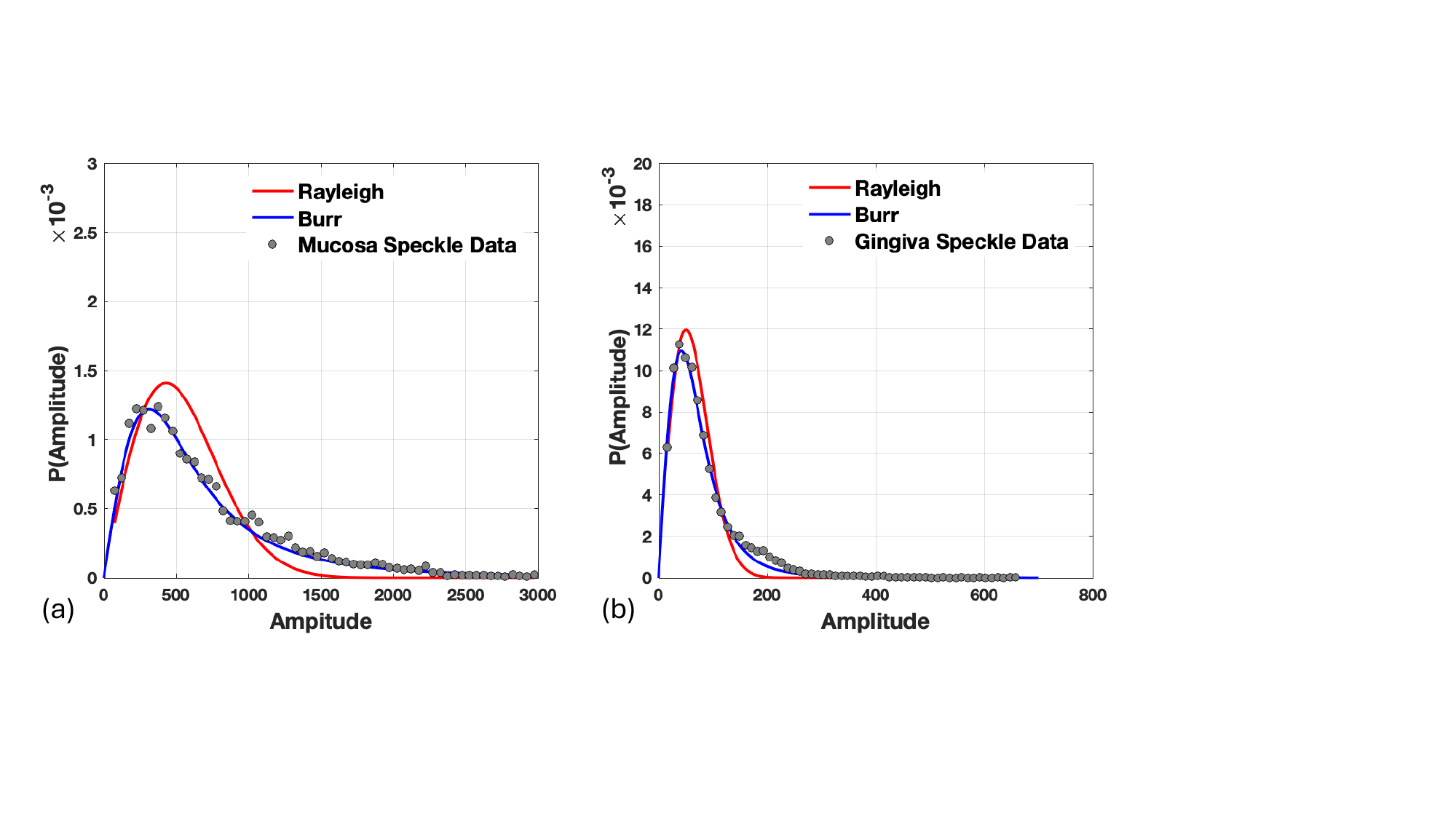}% change 'example-image' to the filename of your image
\caption{Probability distribution of US speckle data fitted to the Rayleigh (red curves) and Burr (blue curves) distributions for \textbf{(a)} alveolar mucosa (Rayleigh fit $R-squared= 0.82$), and \textbf{(b)} gingiva (Rayleigh fit $R-squared=0.95$).}
\label{PDF}
\end{figure}
\end{itemize}
\subsection*{{Limitations and Future Direction}}
In this study, the histology assessment at baseline was based on one sample block of tissues using two different staining methods. In future, we plan to include multiple histology cases and perform statistical and quantitative histology analysis. This will allow a quantitative correlation between histology and QUS. We will investigate histology images through analyzing differences in color densities in gingiva and alveolar mucosa to understand if there is a quantitative correlation between histology findings and QUS parameters in a longitudinal inflammation study. Limitations of this study in terms of the experimental design includes number of porcine for the study and the (still) relatively low resolution of the ultrasound imaging system for scanning small-sized oral tissues. Also, performing QUS analysis on clinical ultrasound data from human subjects should be explored. From the QUS analysis standpoint, future studies should also focus on finding alternate statistical estimators based on higher orders or rational moments of echo envelopes for estimating QUS model parameters and comparing different estimators. Also, the potential effect of WS on the separation accuracy of periodontal tissues could be explored in future. Eventually, future studies should aim at employing other QUS analyses for multi-parametric periodontal diseases classification as well as tissue segmentation. 
%%%%%%%%%%% Conclusions
\section*{Conclusions}
\label{Conclusions}
This study is among early investigations in the literature into applications of QUS approaches for periodontal soft tissue characterizations and serves as a crucial preliminary step with promising results towards employing QUS as an additional diagnostic tool for disease assessment in periodontics. In this study, characterization of periodontal soft tissues (alveolar mucosa and gingiva) was investigated in an \textit{in vivo} swine model using a QUS approach based on the Burr and Nakagami models for speckle statistics. The results showed that the Burr parameters (power-law parameter $b$ and the scale factor $l$) and also Nakagami parameters (shape parameter $m$ and scale factor $\Omega$) have potentials to distinguish clinically significant tissue types. This study demonstrated that the Burr power-law parameter and Nakagami shape parameter were significantly higher in gingiva compared to alveolar mucosa while the Burr scale factor and the Nakagami scale factor were significantly lower in gingiva. The QUS findings were hypothesized to be aligned with qualitative assessments of histology using Masson’s Trichrome and H\&E staining techniques. The two tissue types were classified in 2D parameter spaces using the QUS parameters from the Burr and Nakagami models which yielded a separation accuracy of 93.51\% and 90.91\%, respectively. The classification of the two tissue types using parameters from the two models in 4D resulted in a classification accuracy of 92.21\%. Further studies should assess the effect of disease conditions such as oral soft tissue inflammation on QUS parameters. Our results indicate that QUS could potentially become an augmented tool in periodontics, as an objective, quantitative and noninvasive technique for disease diagnosis, longitudinal monitoring of healing, and feedback for indicated interventions.
\section*{Contributions}
\label{Contri}
\noindent \textit{Daria Poul}: Conceptualization, Methodology, Histology Imaging, Formal Analyses and Software, Validations and Visualization, Writing Original Draft and Revision;\newline
\textit{Ankita Samal, Amanda Rodriguez Betancourt, Carole Quesada}: Animal Experiment, US Scan Data Acquisition, Review and Editing;\newline
\textit{Hsun-Liang Chan, Oliver D. Kripfgans}: Conceptualization, Methodology, Animal Experiment, US Scan Data Acquisition, Funding Acquisition, Review and Editing.

\section*{Conflicts of Interest}
All authors certify that they have no conflicts of interest to report.
%%%%%%%%%%% ACKNOWLEDGEMENTS
\section*{Acknowledgements}
\label{Ack}
The authors acknowledge grant funding from the National Institute of Dental and Craniofacial Research (Grant number: 1-R21-DE-029005).
Also, the authors extend their acknowledgements to Dr Mario Fabiilli from the Department of Radiology, University of Michigan for the loan of the microscopy imaging system.
\section*{Data Availability Statement}
Inquiries regarding the availability of data supporting the findings of this study should be directed to the corresponding authors.
%%%%%%%%%%% REFERENCES
%% REFERENCE FORMATTING INSTRUCTIONS

%% All bibliography information should be included using a 'thebibliography' environment.  Most authors will find it easiest to create a .bbl file using the commands \bibliographystyle{} and \bibliography{} and then copy and paste the contents of the .bbl file into the .tex file below, but before the figure captions section.  Examples for using the \bibliographystyle and \bibliography commands are listed below.  

%% Do not remove the page break here.
\pagebreak

%% References with bibTeX database, use this to create a .bbl file
%\bibliographystyle{UMB-elsarticle-harvbib}
%\bibliography{FILENAME_OF_YOUR_BIBTEX_DATABASE}

%% References copied and pasted from the .bbl file.  Copy and paste over the following two lines.  When using a bibTeX database to create a .bbl file, comment out the following two lines.
% \bibliography{EndNoteExport}  % daria

%%%%%%%%%%% FIGURE CAPTIONS

%% Include only the figure captions here (not the figures).  Figures are uploaded separately in the online Elsevier Editorial Submission process.

%% Do not remove the page break here.
\pagebreak

\section*{Figure Captions}

\begin{description}
\item[Figure 1:]  Anatomical structure of swine periodontal tissues. Several types of fibers within
the gingival tissues provide it with mechanical support during mastication (chewing).
\item[Figure 2:] \textbf{(a)} and \textbf{(b)} High-frequency toothbrush-sized ultrasound transducer for the intraoral scan, \textbf{(c)} the transducer positioning for mid-facial imaging of a molar tooth within the transverse plane in a swine, \textbf{(d)} zoomed-in view of the transducer with the standoff gel pad, \textbf{(e)} an illustration of oral soft tissues anatomy as a general reference for understanding B-scan structure in \textbf{(f)} for a swine case. Important anatomical structures of periodontal hard and soft tissues are annotated in both \textbf{(e)} and \textbf{(f)}.
\item[Figure 3:]  Burr parametric imaging for varying window size (WS in multiples of wavelength)
superimposed on the associated B-mode phantom image. In each pair of parametric images, the left image shows Burr $b$ and the right one shows Burr $l$. All axes are shown in
millimeters.
\item[Figure 4:] Nakagami parametric imaging for varying window size (same phantom study as shown in \textbf{Figure \ref{cirs burr}}). In each
pair, the left image shows Nakagami $m$ and the right one shows Nakagami $\Omega$ parameter local estimations. All axes are shown in millimeters.
\item[Figure 5:] Boxplots showing the effect of WS on the Burr (top row), and the Nakagami (bottom row) parameters. Results in boxplots are from four different imaging planes within an ROI size of 5 mm by 2.5 mm (shown in \textbf{Figure \ref{cirs burr}} and \textbf{Figure \ref{cirs naka}}).
\item[Figure 6:] Parametric imaging of the Burr and Nakagami parameters in periodontal soft tissues in a swine scan using the WS of 10 wavelengths, for maxilla left first molar tooth. \textbf{(a)} reference B-scan, \textbf{(b)} Burr power-law parameter $b$ and \textbf{(c)} Burr scale factor $l$, \textbf{(d)} Nakagami shape parameter $m$ and \textbf{(e)} Nakagami scale factor $\Omega$.
\item[Figure 7:] Parametric imaging of the Burr parameters in the gingiva (top row) and the alveolar mucosa (bottom row) in a swine scan using the WS of 10 wavelengths.
\textbf{(a)} and \textbf{(c)}: Burr $b$, \textbf{(b)} and \textbf{(d)}: Burr $l$. The reference B-scan in shown in \textbf{(e)}.
\item[Figure 8:] Parametric imaging of the Nakagami parameters in the gingiva (top
row) and the alveolar mucosa (bottom row) in a swine scan using the WS of 10 wavelengths.
\textbf{(a)} and \textbf{(c)}: Nakagami $m$, \textbf{(b)} and \textbf{(d)}: Nakagami $\Omega$. The reference B-scan in shown in
\textbf{(e)}
\item[Figure 9:] Burr parameters for classification of gingiva vs. alveolar mucosa, with boxplots summarizing the average estimations within ROIs in swine cases. \textbf{(a)} Burr $b$, \textbf{(b)} Burr $l$. In each boxplot, the blue line shows the statistical median for the population and the
pentagon symbols represent the statistical mean of the estimations.

\item[Figure 10:] Nakagami parameters for classification of gingiva vs. alveolar mu-
cosa, with boxplots summarizing the average estimations within ROIs in swine cases. \textbf{(a)}
Nakagami $m$, \textbf{(b)} Nakagami $\Omega$. In each boxplot, the blue line shows the statistical median for the population and the
pentagon symbols represent the statistical mean of the estimations.
\item[Figure 11:] \textbf{(a)} Histology image using Masson’s Trichrome stain (4x magnification). \textbf{(b)} and \textbf{(c)} are enlarged views comparing gingival and alveolar mucosal regions, respectively (20x magnification). Swine oral site: left mandibular.
\item[Figure 12:] \textbf{(a)} Histology image using  H\&E stain (4x magnification) at the same tissue section as shown in \textbf{Figure \ref{Trichrom}}. \textbf{(b)} and \textbf{(c)} are enlarged views comparing gingival and alveolar mucosal regions, respectively (20x magnification). Swine oral site: left mandibular.
\item[Figure 13:] 2D classifications of gingiva vs. alveolar mucosa in swine cases using
\textbf{(a)} the Burr model (Burr $b$ vs. Burr $l$), and \textbf{(b)} the Nakagami model (Nakagami $m$ vs.
Nakagami $\Omega$). Black lines show linear boundaries between two classes of tissues. Blue
diamond symbols: estimations for gingiva, orange circle symbols: estimations for mucosa.
\item[Figure 14:] 3D Classification of gingiva vs. alveolar mucosa using combined
parameters of the Burr and Nakagami models, represented in the principal component
space using the first three principal components of PC 1, PC 2, and PC 3.
\item[Figure 15:] Probability distribution of US speckle data fitted to the Rayleigh (red curves) and Burr (blue curves) distributions for \textbf{(a)} alveolar mucosa (Rayleigh fit $R-squared= 0.82$), and \textbf{(b)} gingiva (Rayleigh fit $R-squared=0.95$).

\end{description}

%% Do not remove the page break here.
\pagebreak
\end{document}